\documentclass[twocolumn]{aastex631}

\usepackage{enumitem}
\usepackage{subfigure}

\begin{document}

\title{VHE FSRQs with Fermi-LAT: VHE and even brighter states in high-z FSRQs due to an HBL-like component?}
\author[0009-0008-7119-9770]{Megha}
\affiliation{Department of Physical Sciences, Indian Institute of Science Education and Research (IISER) Mohali \\
Knowledge City, Sector 81, SAS Nagar, PO Manauli 140306, India}
\email{ph23030@iisermohali.ac.in}

\author{Pankaj Kushwaha}
\affiliation{Department of Physical Sciences, Indian Institute of Science Education and Research (IISER) Mohali \\
Knowledge City, Sector 81, SAS Nagar, PO Manauli 140306, India}

\collaboration{20}{(AAS Journals Data Editors)}



\begin{abstract}

Very high-energy (VHE) detected flat-spectrum radio quasars (FSRQs) are relatively few despite being the most persistent bright MeV-GeV sources. Focusing on VHE emission, we investigated the spectral and temporal properties of VHE-detected FSRQs using 14-year Fermi-LAT data. All are highly variable (flux-amplitude$>$100) with VHE detection associated with brighter flux states and relatively harder spectra. Above a flux limit, flux anti-correlates with spectral index, exhibiting a bluer-when-brighter trend. The low-flux state spectral energy distributions (SEDs) for all resembles a power-law, while high-flux and VHE-associated states resemble a log-parabola, accompanied by an almost nil (PKS0736+017, PKS1510-089) to marginal (4C+21.35, 3C279) to significant (B21420+32, TON0599, PKS1441+25, S30218+35, PKS0346-27, OP313) MeV-GeV peak-upshift -- more prominent in high-redshift sources. For no/marginal peak-upshift, the VHE emission is consistent with external Comptonization of infrared photons (EC-IR) driven primarily by a power-law continuation of the particle spectrum to higher energies. For those with a significant MeV-GeV peak-upshift, PKS0346-27 and OP313 shows peak-upshift in the synchrotron spectrum, and thus VHE is EC-IR origin, while for others without synchrotron-peak upshift, we attribute the VHE to an HBL-like component with a Compton-Dominance (CD) like FSRQs, with VHE driven primarily by particle spectrum continuation. In some, even high-state SEDs seem to require an HBL-like component. Thus, VHE activities in FSRQs mainly result from particle spectrum continuation, aided by spectral transition or a new HBL-like component with FSRQ-like CD. Such spectral changes naturally brighten the GeV-VHE flux, overcoming extragalactic background light absorption without requiring extraordinary brightening under the traditional EC-IR scenario than normally exhibited.

\end{abstract}




\section{Introduction} \label{sec:intro}
Blazar is an active galactic nuclei (AGN) subclass designated radio-loud that harbors large-scale, powerful bipolar relativistic jets directed almost along the line joining the source and the Earth \citep[$\leq 15^\circ$;][]{UrryPadovini1995}. 
It comprises traditionally classified BL Lacertae objects (BL Lacs) and flat-spectrum radio quasars (FSRQs). The traditional classification is based on the strength of optical emission lines with respect to the underlying continuum, which is measured in terms of the equivalent width (EW). In this scheme, BL Lacs have EW $\leq5~A^\circ$, while FSRQs have strong broad emission lines \citep{Blandford_Rees1978}. The observational traits of blazars are a rapidly variable non-thermal continuum spanning all the accessible EM spectrum from radio up to GeV/TeV gamma-ray energies, high and variable polarization at radio, optical, and even at X-ray in many \citep{Faan2008,Itoh2016,Middei2023}, superluminal features in radio images, and intraday variability \citep{Wagner1995_intradayVariability}.

The spectral hallmark characteristic of a blazar's broadband continuum is its broad bimodal spectral energy distribution \citep[SED;][]{blazarClassification,SEDdoubleHump}. The lower-energy component extends from radio up to X-rays, peaking at near-infrared (NIR) to X-rays, while the high-energy component extends from X-rays up to TeV, peaking in the MeV-GeV energy bands. One interesting aspect of this bimodal shape is the frequency at which the low-energy component peaks is exceptionally stable \citep[with few exceptions, e.g., Mkn 501, PKS 1441+25;][]{Pian1998,Ahnen2015_pks1441} despite huge flux variations. Further, the location of this peak anticorrelates with the source luminosity and shows a sequence classification as low-synchrotron-peaked (LSP), intermediate-synchrotron-peaked (ISP), and high-synchrotron-peaked (HSP) \citep{SEDdoubleHump}.

A strong and variable linear polarization at radio, optical, and even X-rays in some and the non-thermal nature of spectra strongly suggest that the low-energy hump is synchrotron emission from the relativistic electrons gyrating in the magnetic field inside the jet. The origin of the high-energy hump is one of the highly debated open issues, and the plausible emission scenarios involve relativistic leptons or hadrons \citep[primarily protons;][]{Bottcher2013}. The hadronic scenario attributes the high-energy hump to either individually or a combination of proton synchrotron radiation and/or proton-proton, proton-photon initiated cascades leading to pion production \citep{Mannheim1992,Mucke2001}. The claim of association of neutrinos from a few blazars supports hadronic processes, while SED modeling suggests an insignificant contribution to MeV-GeV emission, indicating that MeV-GeV emission is primarily due to leptonic processes \citep{Gao,Liodakis_2020,Acharyya_2023}. 

The leptonic channel explains the high-energy emission via inverse Compton (IC) scattering \citep{ICscattering,Ghisellini_tavecchio2009} since there is an abundance of a wide variety of prominent soft-photon fields, like the broad-line region \citep[BLR;][]{EC_BLR}, IR photons from the dusty torus \citep{EC_IR}, accretion disc \citep[AD;][]{EC_AD}, or cosmic microwave background (CMB) photons \citep{EC_CMBR} apart from the synchrotron photons themselves \citep[SSC model;][]{SSCmodel}.

FSRQs are strong gamma-ray emitters, with the total radiative budget dominated entirely by the MeV-GeV emission, yet they appear to be weak VHE emitters. So far, only 10 out of 750 \textit{Fermi}-detected FSRQs \citep{ballet2024_4fglCatA} have been detected at VHE\footnote{\url{http://tevcat2.uchicago.edu/}}. In the leptonic scenario, FSRQs' MeV-GeV emission is argued to be predominantly due to IC scattering of BLR (IC-BLR) and IR (IC-IR) photon fields. For VHE, IC-BLR is irrelevant due to the Klein-Nishina effect, restricting the output to 20-30 GeV. Further, BLR photons provide strong opacity to the VHE photons through photo pair production, making the BLR opaque to the VHE and implying the emission region to be beyond the BLR \citep[e.g.][]{bottcher2016}. 

The location of the emission region puts a constraint on the predominant soft-photon field for IC scattering. However, observationally, we observe a smooth spectrum that extends well beyond 100 GeV, implying that the very high energy (VHE; $E>100~GeV$) is of non-IC-BLR origin.
A harder VHE spectrum is observed, which suggests that IC scattering must take place in the Thomson regime. Thus, for VHE emission, a softer photon field is required, and IR photons from the dusty torus are one of the natural and generally favored plausibilities.

In this work, we study the nine\footnote{The 10th VHE FSRQ (OP313) was reported in December 2023.} VHE FSRQs (see Table \ref{tab: 4.1}) to investigate the role of BLR and IR in gamma-ray emission, thus indirectly probing the location of the emission region as well as one-zone or multi-zone emission scenarios. To achieve this, we conducted a comprehensive analysis of the spectral and temporal behavior exhibited by VHE FSRQs during phases of low, high, and VHE gamma-ray activities. The work is organized into five sections. In \S\ref{sec:DataReduction}, we discussed the {\it Fermi} LAT data reduction and the gamma-ray lightcurve and SED analysis procedure. In \S\ref{sec:Analysis}, we present the analysis methods used along with our results, followed by a detailed discussion on plausible emission scenarios and the particle spectrum in \S\ref{sec:Discussion}. We conclude in \S\ref{sec:summary}.

\begin{table}
    \caption{Basic details of the nine VHE-FSRQs considered in this work.} 
    \centering
    \begin{tabular}{|l|l|l|l|c|}
    \hline
    Name         & RA          & DEC          & z$^*$ & $F_R$ \\
    \hline
    \hline
    PKS 0736+017 & 07 39 17.0  & +01 36 12    & 0.18941$^1$  & 168 \\
    PKS 1510-089 & 15 12 52.2  & -09 06 21.6  & 0.361$^2$  & 902 \\
    4C +21.35   & 12 24 54.4  & +21 22 46    & 0.432$^3$ & 314 \\
    3C 279      & 12 56 11.1  & -05 47 22    & 0.5362$^4$  & 144 \\
    B2 1420+32  & 14 22 30.38 & +32 23 10.44 & 0.682$^5$  & 140  \\
    TON 0599    & 11 59 31.8  & +29 14 44    & 0.7247$^{6}$  & 106 \\
    PKS 1441+25  & 14 43 56.9  & +25 01 44    & 0.939$^{7}$  & 178 \\
    S3 0218+35   & 02 21 05.5  & +35 56 14    & 0.954$^{8}$  & 626 \\
    PKS 0346-27  & 03 48 38    & -27 49 14    & 0.991$^{9}$  & 134  \\
    OP 313$^{**}$ & 13 10 28.6638  & +32 20 43.783  & 0.9973$^{10}$ & -- \\
    \hline
    \end{tabular}
    \begin{tablenotes}
        \item  $\rm F_R = maximum-flux/minimum-flux$
    \end{tablenotes}  
    $^1$ \citep{Ho2009PKS0736_redshift}
    $^2$ \citep{Tanner1996PKS1510_redshift} 
    $^3$ \citep{Osterbrock1987_4C_redshift}
    $^4$ \citep{Burbidge1965_3C_redshift}
    $^{5}$ \citep{Hewett2010TON_B2_redshift}
    $^{6}$ \citep{Hewett2010TON_B2_redshift}
    $^{7}$ \citep{Shaw2012PKS1441_redshift}
    $^{8}$ \citep{Cohen2000S3_redshift} 
    $^{9}$ \citep{White1988PKS0346_redshift}
    $^{10}$ \citep{2010AJ_op313_redshift} \\
    $^{**}$ We did not consider this source as it was detected at VHE \citep{2024_op313_vhe} after the time duration taken by us.
    \label{tab: 4.1}
\end{table}


\section{FERMI LAT Data reduction} \label{sec:DataReduction}
The {\it Fermi} observatory was launched on 11 June 2008 with two astronomical payloads onboard: the Gamma-ray Burst Monitor (GBM), primarily for studying gamma-ray bursts, and the Large Area Telescope (LAT), a more sensitive MeV-GeV instrument for surveys. LAT works in the energy range of 20 MeV to $>$300 GeV with a field of view (FoV) of 2.4 sr. Most of the time, LAT works in 'scanning mode', in which it scans the whole sky every 3 hours \citep{FermiLAT}. {\it Fermi} does not take data while moving over the South Atlantic Anomaly (SAA). On 16 March 2018, a Solar Array Drive Assembly (SADA)\footnote{\url{https://fermi.gsfc.nasa.gov/ssc/observations/types/post_anomaly/}} anomaly occurred, resulting in a lower exposure for high and low declination sources. 

For our sources of interest listed in Table \ref{tab: 4.1} (ascending redshift), we used LAT data between August 2008 and August 2022 and followed the standard data reduction procedures. We performed binned likelihood analysis using {\it Fermipy} version \textit{'v1.2'}. For our analysis, events belonging to the Source class (evclass=128, evtype=3) and energies between 100 MeV and 800 GeV were used, and a maximum zenith angle cutoff of $90^{\circ}$ was applied to reduce the contamination from the Earth's limb. The analysis was performed with the instrument response function P8R3\_SOURCE\_V3, the galactic diffuse emission model (gll\_iem\_v07.fits), and the isotropic background model (iso\_P8R3\_SOURCE\_V3\_v1.txt) employing the binned maximum likelihood method implemented in the Science tool \textit{gtlike}. The VHE FSRQs are modeled using the default model spectrum as mentioned in the {\it Fermi} 4FGL catalog \citep{ballet2024_4fglCatA,Abdollahi2022_4fglCatB}, which is either a power law (PL) or a log parabola (LP) with the functional form, 
\begin{equation}
    PL:~~~ \frac{dN}{dE} = N\left ( \frac{E}{E_o} \right )^{-\Gamma }
\end{equation}
\begin{equation}
    LP:~~~ \frac{dN}{dE} = N\left ( \frac{E}{E_o} \right )^{-(\alpha +\beta log(E/E_o))}
\end{equation}
where $dN/dE$ is the differential photon flux in units of $ph~cm^{-2}~s^{-1}~MeV^{-1}$, N is the normalization factor, $E_o$ is the scale factor, $\alpha$ and $\Gamma$ are the photon indices, and $\beta$ is the curvature parameter. 

Firstly, we extracted monthly binned lightcurves using the \textit{lightcurve} module of {\it Fermipy}. For this, events with energies between 0.1 and 800 GeV were selected from an acceptance cone of $15^{\circ}$ centered at the location of the source of interest. The model parameters of 4FGL sources within a radius of $10^{\circ}$ from the source of interest were left free to vary. Sources with test statistics (TS) $>$0 and $n_{pred}>$30 were chosen for the analysis.

For extracting the SEDs, we used the \textit{SED} module of \textit{Fermipy} in the energy range of 0.1-800 GeV. The number of energy bins was set to 12, and \textit{use\_local\_index} was set to True. SEDs have been generated for the low and high gamma-ray flux states, the VHE activity period, and the entire duration considered in this work.

We used the \textit{Fermipy} flux sensitivity, which provides the value of the minimum photon flux that LAT can detect at a given energy. To further assess the reliability of the extracted SED data points, we calculated sensitivity using the \textit{sensitivity} tool of {\it Fermipy}. The galactic diffuse model \textit{gll\_iem\_v07.fits} and event class \textit{P8R2\_SOURCE\_V6} are used along with a TS threshold of 10.0 and a minimum number of counts of 5.0.

\begin{figure*}[ht!]
\includegraphics[scale=0.4]{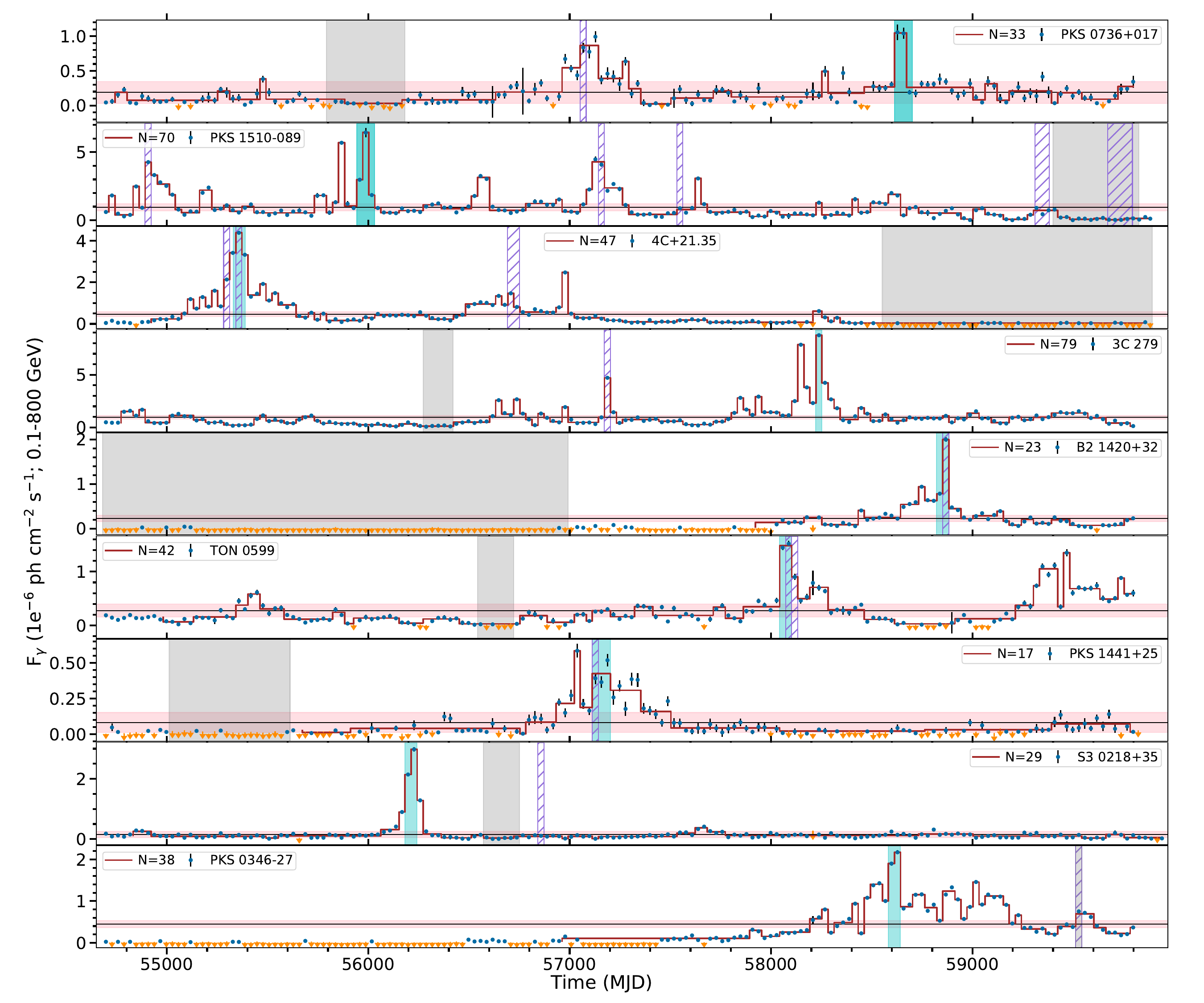}
\caption{Monthly binned light curve of the 9 VHE FSRQs (Table \ref{tab: 4.1}) for the duration of 5 August 2008 to August 2022. The {\it Fermi} data is marked with blue points along with the error bars in black. Orange arrows mark the upper limits (TS$<9.0$). The Bayesian block is shown in a brown-colored curve, and the number of bins is given by N. The dashed purple region marks the duration coincident with the VHE detection for which VHE SEDs have been extracted. The horizontal black line is the mean flux with a 3-sigma red color band around the mean. The grey shaded regions mark the duration used for the low flux state SED extraction (ref. \S\ref{sec:lowHighstate}), and the cyan region is used for the high flux state SED. References for VHE detection are: PKS 0736+017 \citep{HESS2020_0736_VHE}, PKS 1510-089 \citep{HESS2013_1510_VHE_2009,2017_1510_VHE_2016,Aharonian_2023}, 4C+21.35 \citep{MAGIC2011_4c_VHE_2010,2015_4c_VHE_2014}, 3C 279 \citep{HESS2019_3C279_2015_VHE}, B2 1420+32 \citep{MAGIC2021_B2_2020_VHE}, TON 0599 \citep{2017ATel_TON_2017}, PKS 1441+25 \citep{Ahnen2015_pks1441}, S3 0218+35 \citep{Ahnen2016_S3_HBLtpe}, PKS 0346-27 \citep{2021ATel_0346_VHE}.}
\label{fig:LC}
\end{figure*}


\section{Data analysis and Results}  \label{sec:Analysis}
The monthly photon flux lightcurve extracted by the procedure outlined in \S\ref{sec:DataReduction} for the 9 VHE FSRQs is shown in Figure \ref{fig:LC}. Data points with TS$<9.0$ are plotted as upper limits. The mean of the photon flux and the 3-$\sigma$ band is marked in the light curve with $\sigma$ defined as the mean of the photon flux error. We used a Bayesian block \citep{Scargle2013} to mark and identify the photon flux changes (marked by the red solid line). It is a very adaptable and robust statistical method enabling effective identification of significant flux changes. This method segregates the light curve into optimal bins given by N, which is also mentioned in the figure. The low and high flux states used for SED extraction (\S\ref{sec:lowHighstate}) are marked with gray and cyan shaded regions, respectively. Episodes of VHE detection by ground-based observatories (MAGIC\footnote{\url{https://magic.mpp.mpg.de/}}, HESS\footnote{\url{https://www.mpi-hd.mpg.de/hfm/HESS/}}, and VERITAS\footnote{\url{https://veritas.sao.arizona.edu/}}) are marked by a purple dashed region. 

We found that all the sources are highly variable, with a variation of $>100$ times between the maximum and minimum detected photon flux. This ratio is listed in the last column of Table \ref{tab: 4.1}, and based on this ratio, the maximum amplitude variation is exhibited by PKS 1510-089 and the minimum by TON 0599 on the monthly time scale. Among these sources, B2 1420+32 and PKS 0346-27 are the ones that remain inactive and below the sensitivity limit for most of the duration, with the percentage of detection bins being 41\% and 51\%, respectively.

\subsection{Histogram} \label{sec:histogram}

\begin{figure}
    \centering
        \includegraphics[scale=0.35]{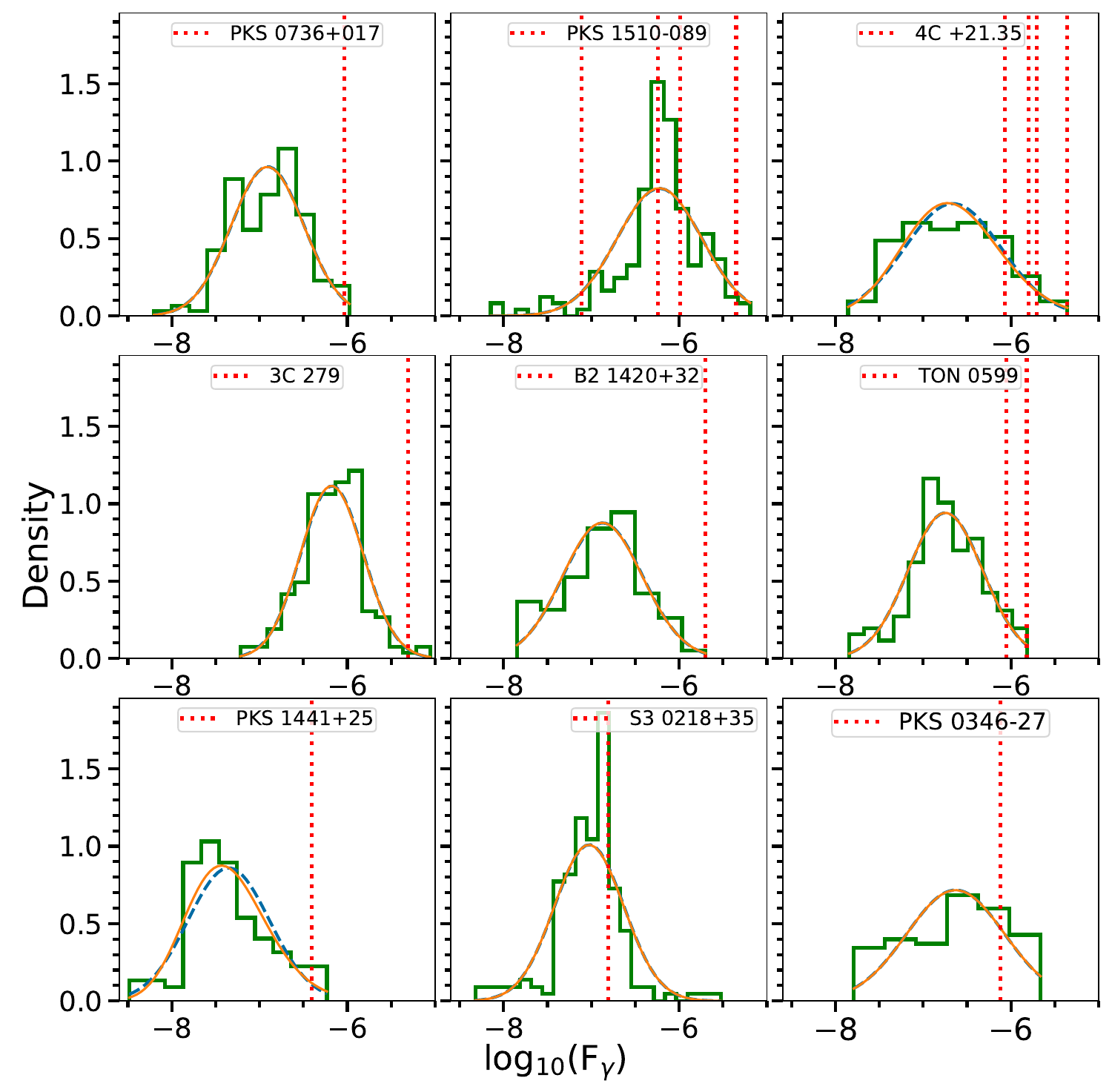}
        \includegraphics[scale=0.35]{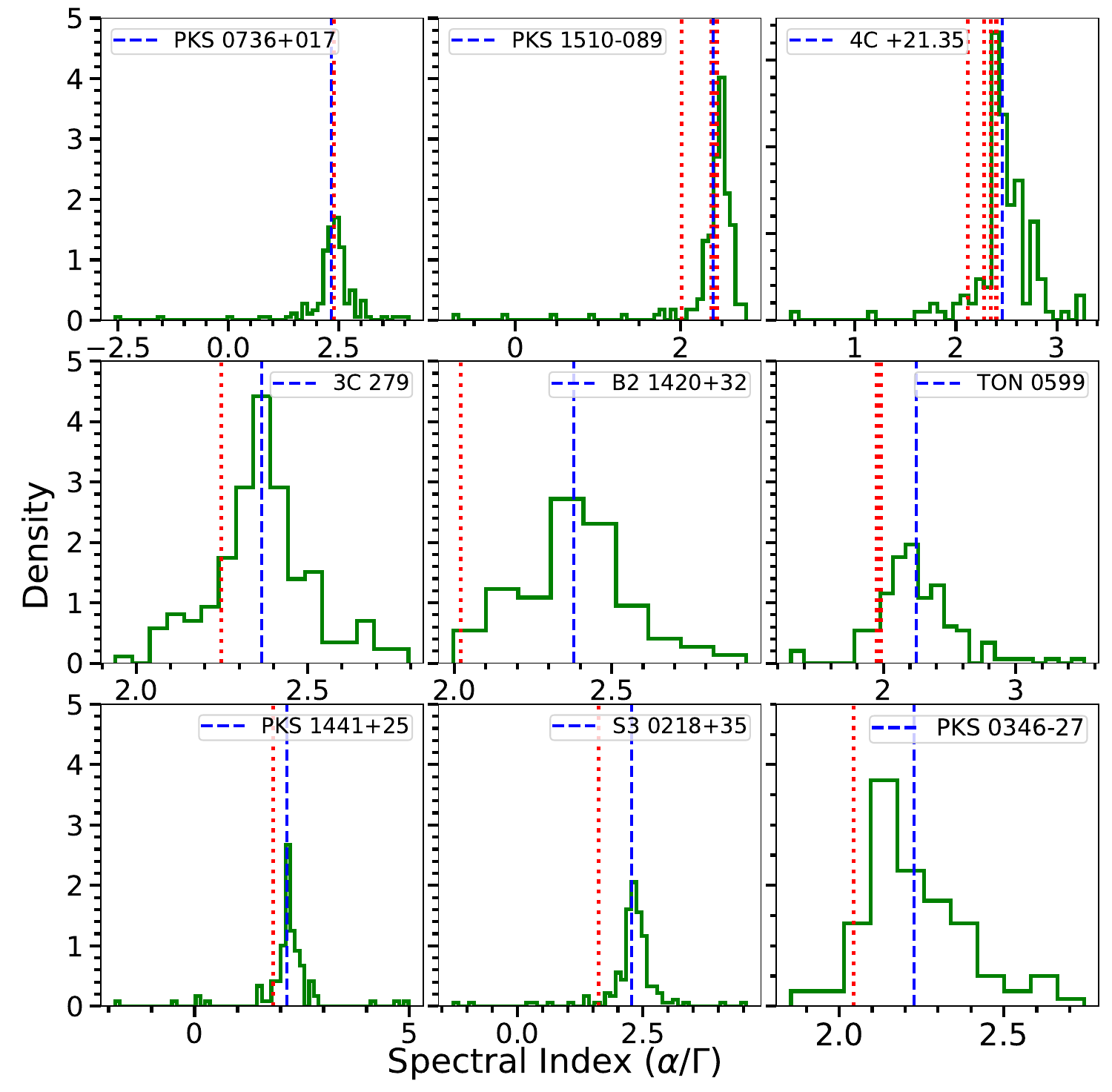}
        \caption{{\bf Top}: Log(flux) histogram plot for all 9 VHE FSRQs with increasing order of redshift from left to right. The solid orange line is a normal fit, and the dashed blue line represents a lognormal fit. The dotted vertical red line represents the value of the log(flux) at the time of the VHE episode. {\bf Bottom: }Spectral index ($\alpha/\Gamma$) histogram plot for all 9 VHE FSRQs with increasing order of redshift from left to right. The dotted red line marks the spectral index at the time of the VHE episode, and the dashed blue line marks the mean spectral index.
        \label{fig:SIhist}}
    \label{fig:enter-label}
\end{figure}


\begin{figure}[ht!]
\includegraphics[scale=0.3]{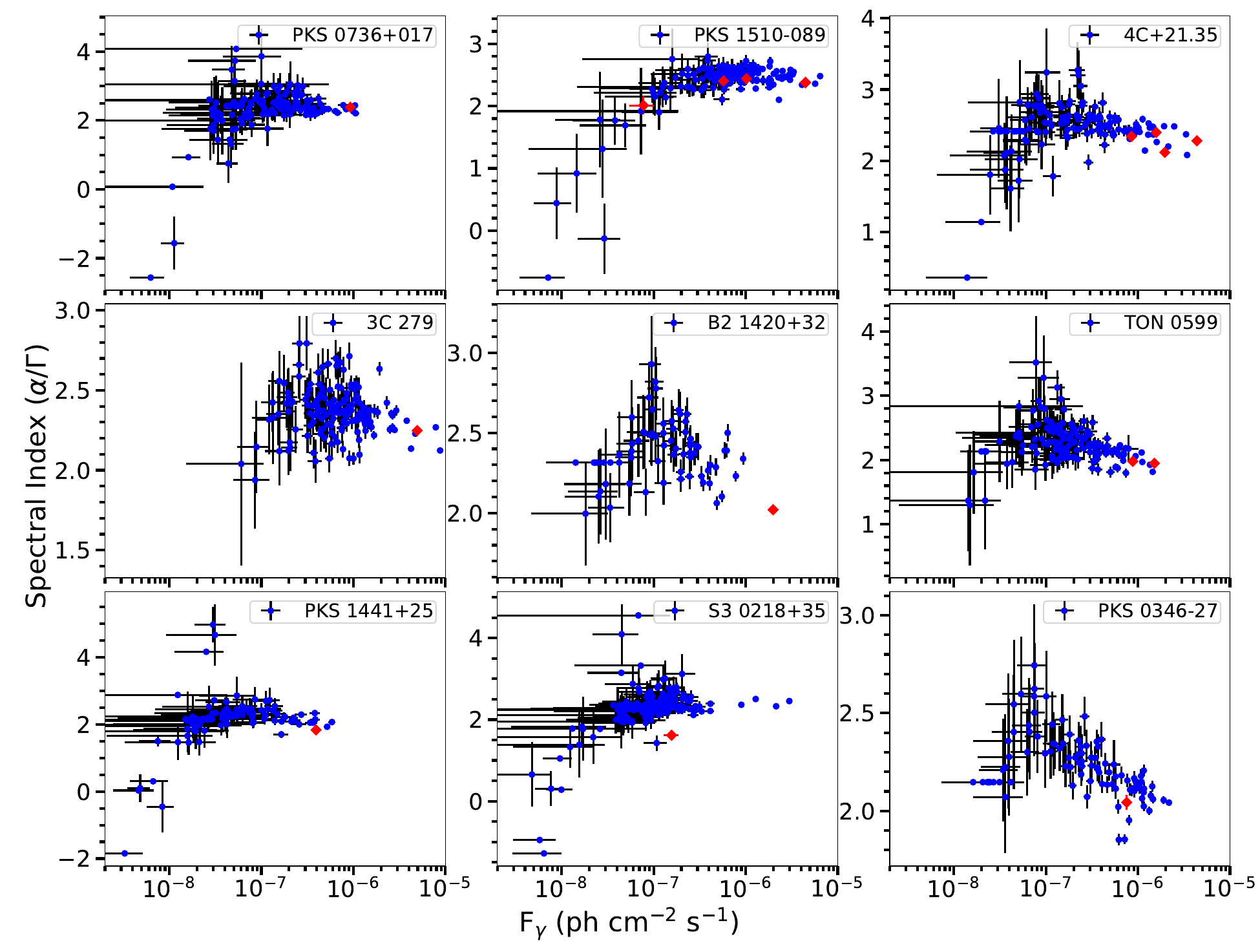}
\caption{Spectral index ($\alpha/\Gamma$) versus flux trend for sources in increasing order of redshift from left to right. Red points mark the MeV-GeV spectral index and photon flux during the time of the VHE episode.
\label{fig:SIvsFlux}}
\end{figure}

To identify the most observed photon flux state and spectral state at MeV-GeV, we have generated histograms for the logarithm\footnote{to the base 10} of photon flux (Figure \ref{fig:enter-label} top) and the spectral index (Figure \ref{fig:enter-label} bottom) for all, considering only the detected points. The time of VHE activity is marked with red dotted lines. Histograms were generated employing the Knuth method \citep{knuth2013optimal}. 

We further tested the histogram of log(photon-flux) with Gaussian and lognormal distributions. We also applied the Kolmogorov-Smirnov (KS) test to assess the goodness-of-fit of the observed log(photon-flux) histogram with both models. The null hypothesis is that there is no significant difference between the data and the normal distribution. For the normal distribution, we obtained the statistics and p-values, respectively, from 0.053 to 0.135 and 0.003 to 0.939. For the lognormal model, we obtained statistics and p-values from 0.053 to 0.100 and 0.061 to 0.933. Based on the p-values, we cannot reject our null hypothesis in general, suggesting that both the normal and lognormal distributions reasonably represent the data well.  
We observe a continuous distribution of photon flux, and the MeV-GeV spectral indices associated with VHE episodes are between 1.62 and 2.44, in general relatively harder during VHE episodes (Figure \ref{fig:enter-label} bottom panel and Table \ref{tab:Params}). 

\subsection{Spectral index versus photon flux plots} \label{sec:SIvsFlux}
We also investigated the relation between spectral index and photon flux, and the corresponding plot is shown in Figure \ref{fig:SIvsFlux}.
The MeV-GeV spectral indices coinciding with the time of VHE activity are represented in red color. Although visual inspection does
not indicate any clear pattern or trend, we found a clear trend of bluer-when-brighter (BWB; spectral hardening during brightening) above some flux limit.
The lack of any discernible pattern as a whole is likely due to detection in a one/two energy bin during low fluxes and, thus, a relatively harder spectral index. The flux limits (in $ph~cm^{-2}~s^{-1}$ units) for different sources are: PKS 1510-089 ($F\gtrsim 1\times 10^{-6}$), 4C+21.35 ($F\gtrsim 1\times 10^{-7}$), TON 0599 ($F\gtrsim 1\times 10^{-7}$), PKS 1441+25 ($F\gtrsim 5\times 10^{-8}$), and PKS 0346-27 ($F\gtrsim 1\times 10^{-7}$), while for PKS 0736+017 and 3C 279, a slight indication of the trend is seen. For B2 1420+32 and S3 0218+35, no trend is seen. It also clearly shows that VHE detections generally occur during the brighter MeV-GeV flux states and have a relatively harder spectral index (refer to Figure \ref{fig:SIhist}).


\subsection{Low and high flux states SED} \label{sec:lowHighstate}
Motivated by the association of VHE detection with brighter MeV-GeV flux states, we extracted the MeV-GeV SEDs during high- (no VHE) and low-flux states and also for the time period coinciding with the VHE episodes for a comparative spectral investigation.
Various researchers have adopted different criteria for defining low- and high-flux states based on the focus of the study. For example, \citet{Meyer_georganopolous2014} adopted a significance-based approach for producing the low-state SED. They have used the method of progressive binning to find the minimum flux based on a chosen TS. \citet{Meyer} introduced a more refined way of identifying the low-flux state using light curves extracted using shorter time bins. However, in a photon-starved EM window, such as gamma rays, identifying a few time bins of low-flux states does not yield sufficient count statistics for spectral investigations. \citet{Costamante} has used a fixed flux value of $10^{-6} ~ph~cm^{-2}~s^{-1} $, a value used by the {\it Fermi}-LAT team to categorize the source as bright\footnote{\url{https://fermi.gsfc.nasa.gov/ssc/data/access/lat/msl_lc/}}.

\begin{figure*}
\centering
\includegraphics[width=18cm, height=7.5cm]{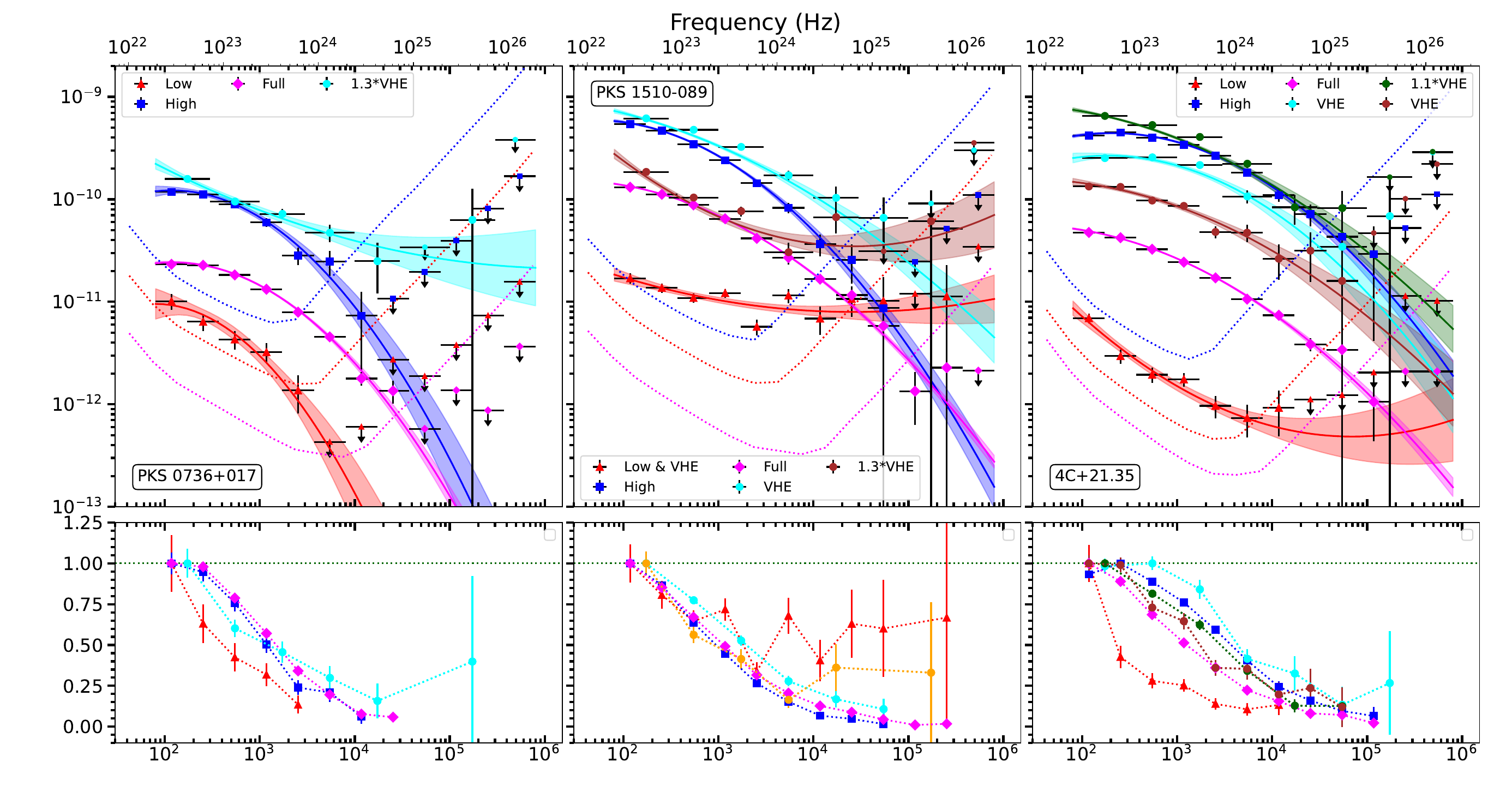}
\includegraphics[width=18
cm, height=7.5cm]{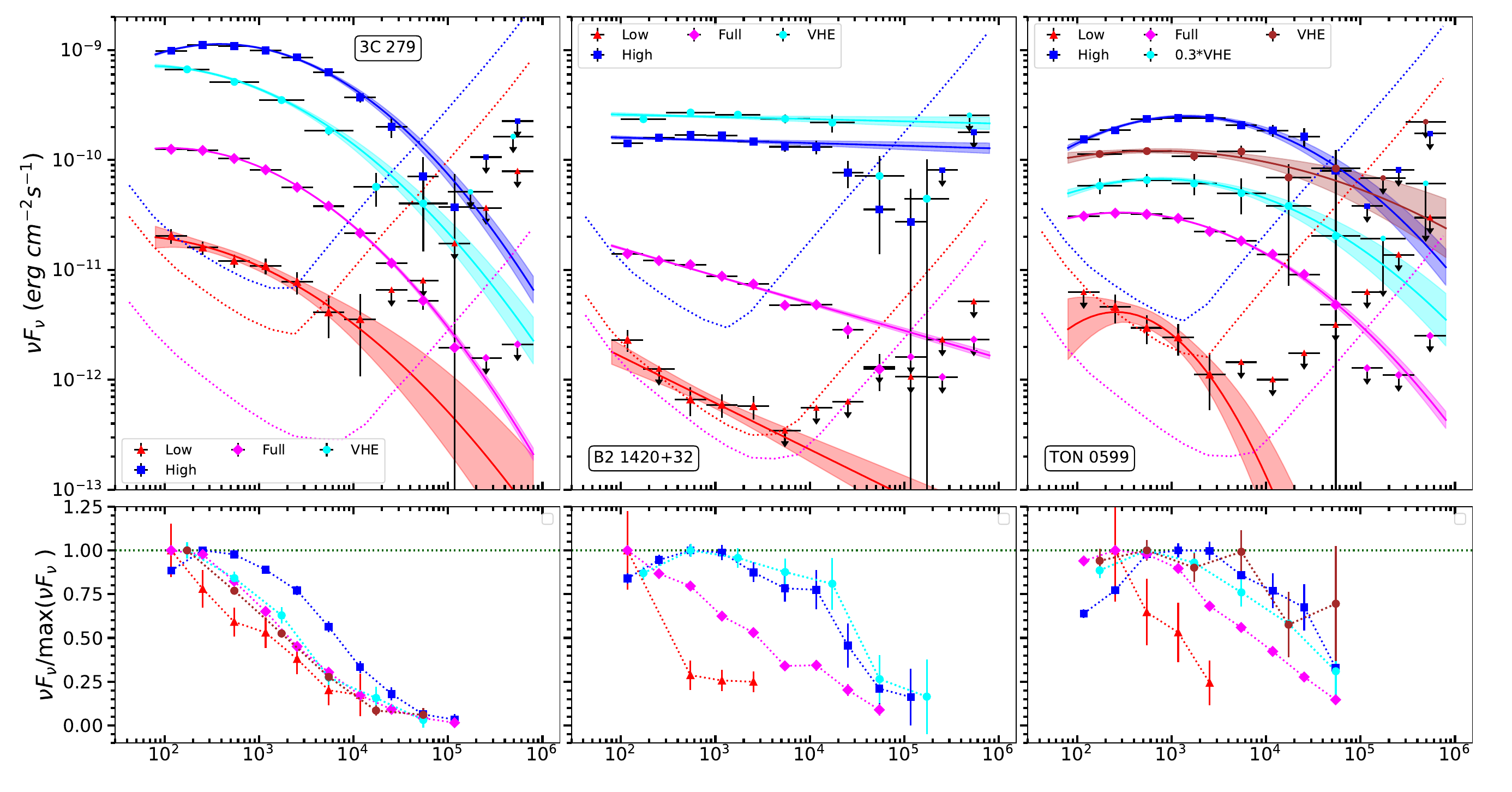}
\includegraphics[width=18cm, height=7.5cm]{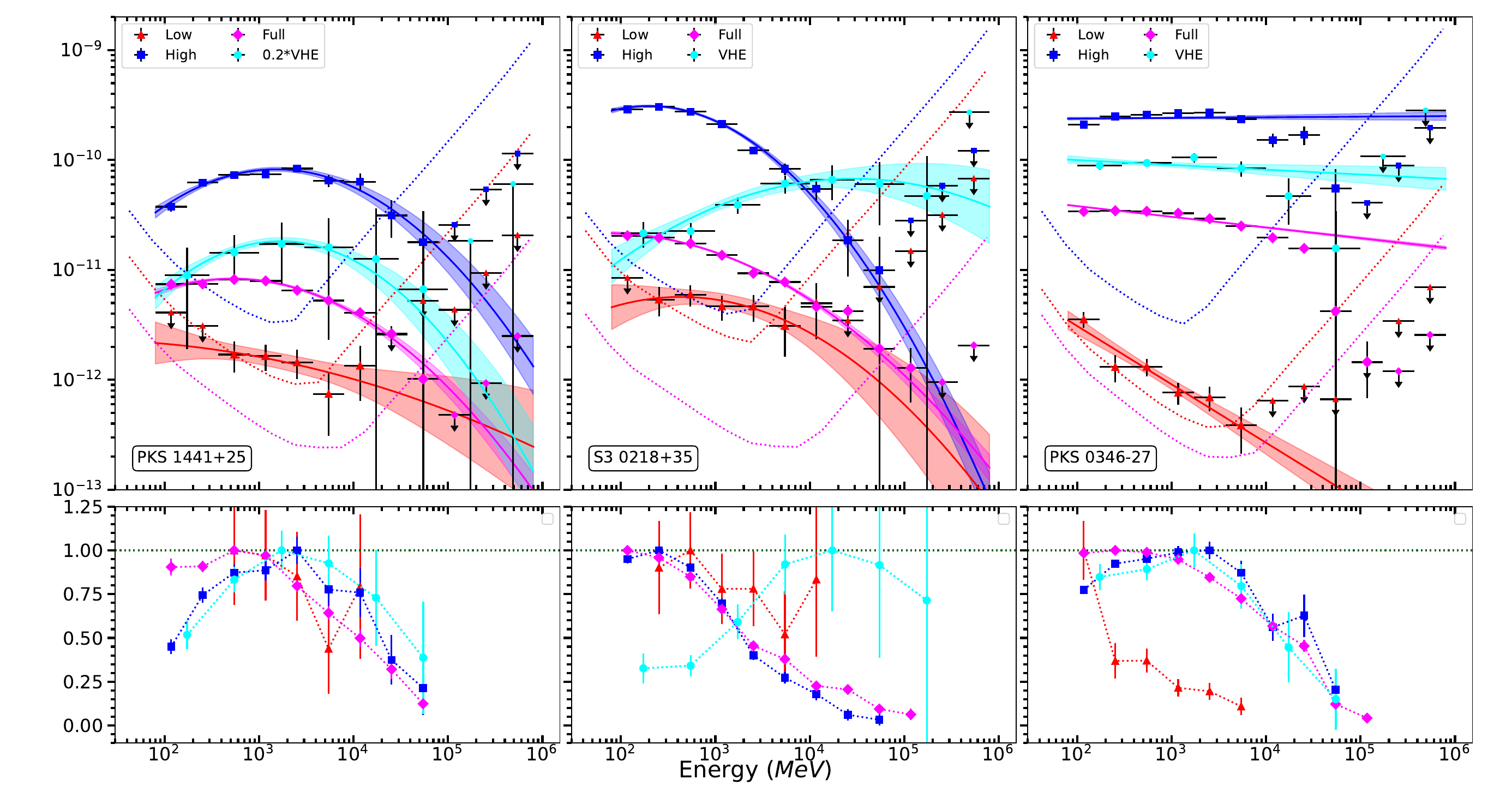}
\caption{MeV-GeV SEDs from LAT for the 9 VHE FSRQs in increasing order of redshift during low (Low) and high (High) flux states, along with the SED for the entire observation period (Full) and VHE detected durations (VHE; see Figure \S\ref{fig:LC}). The black arrows mark the upper limits, with dotted curves showing the Fermi-LAT sensitivity for the corresponding states (\S\ref{sec:DataReduction}). The numerical factor in the `VHE' label is a scaling factor for a clearer representation. The solid curves are the best-fit model (PL/LP; see Table \ref{tab:Params}) with a 1-$\sigma$ shaded region. The lower panel of the SEDs gives the ratio of $\nu F_{\nu}$ to max($\nu F_{\nu}$), a good proxy for measuring peak shifts. The dotted horizontal line marks the value of 1.0.
\label{fig:SEDwithVHE}}
\end{figure*}

Our focus is on comparing the different flux state SEDs, which hence requires a good number of photons. Thus, for bright phases, a shorter duration will suffice, so we have chosen the data corresponding to the highest flux only. For low-flux state SED, a longer duration is considered when the source is in low-flux states and/or mostly upper limits, i.e., below the detection limit. This method is broadly similar to \cite{Meyer_georganopolous2014}, allowing us the extraction of spectra up to and beyond 10 GeV. The SEDs and the best-fit model, along with the 1-$\sigma$ error, are shown in Figure \ref{fig:SEDwithVHE}, and the parameters are given in Table \ref{tab:Params}. Again, data points with $TS<9.0$ are marked as upper limits. Additionally, we have added the LAT sensitivity curve for each SED, shown by the dotted curves in the respective colors.

\begin{table}
  \caption{The best-fit spectral parameters for SEDs corresponding to different flux states (refer to \S\ref{sec:lowHighstate}). The spectral parameters with bars  ($\bar{\alpha},\bar{\beta},\bar{\Gamma}$) are the mean values estimated from the light curve.}
    \centering
    \begin{tabular}{|c|c|c|c|c|c|c|c|c|c|}
    \hline
       Source Name & State & alpha (LP) & beta \\ 
    \hline
    \hline
       PKS 0736+017 & Low & 2.67$\pm$0.13 & 0.18$\pm$0.12 \\ 
        $\bar{\alpha}=2.33\pm0.26$ & High & 2.38$\pm$0.04 & 0.13$\pm$0.04 \\ 
        $\bar{\beta}=0.46\pm0.22$ & Full & 2.33$\pm$0.01 & 0.116$\pm$0.009 \\ 
         & VHE & 2.38$\pm$0.05 & -0.02$\pm$0.03 \\
    \hline
       PKS 1510-089 & Low+VHE & 2.18$\pm$  0.05 & -0.02$\pm$ 0.02 \\ 
        $\bar{\alpha}=2.39\pm0.12$ & High & 2.48$\pm$0.02 & 0.09$\pm$0.01 \\ 
        $\bar{\beta}=0.07\pm0.07$ & Full & 2.398$\pm$0.004 & 0.059$\pm$0.003 \\ 
         & VHE & 2.38$\pm$0.02 & 0.04$\pm$0.01 \\
         & VHE & 2.44$\pm$0.05 & -0.06$\pm$0.23 \\
    \hline
       4C+21.35 & Low & 2.65$\pm$0.08 & -0.06$\pm$ 0.03 \\ 
        $\bar{\alpha}=2.46\pm0.16$ & High & 2.10$\pm$0.01 & 0.079 $\pm$0.008 \\ 
        $\bar{\beta}=0.14\pm0.14$ & Full & 2.303$\pm$ 0.007 & 0.054$\pm$0.004 \\ 
         & VHE & 2.12$\pm$0.05 & 0.08$\pm$0.02 \\
         & VHE & 2.28$\pm$0.02 & 0.04$\pm$0.01 \\
         & VHE & 2.25$\pm$0.03 & 0.04$\pm$0.02 \\
    \hline
       3C 279 & Low & 2.29$\pm$ 0.09 & 0.06$\pm$0.05 \\ 
        $\bar{\alpha}=2.36\pm0.09$ & High & 2.024$\pm$ 0.009 & 0.088$\pm$0.006 \\ 
        $\bar{\beta}=0.09\pm0.07$ & Full & 2.222$\pm$0.004 & 0.082$\pm$0.003 \\ 
         & VHE & 2.24$\pm$0.02 & 0.06$\pm$0.01 \\
    \hline
       TON 0599 & Low & 2.32$\pm$0.26 & 0.26$\pm$0.19 \\ 
        $\bar{\alpha}=2.25\pm0.18$ & High & 1.83$\pm$0.02 & 0.079$\pm$0.009 \\ 
        $\bar{\beta}=0.15\pm0.14$ & Full & 2.071$\pm$0.007 & 0.069$\pm$0.004 \\ 
         & VHE & 1.95$\pm$0.03 & 0.06$\pm$0.02 \\
         & VHE & 1.98$\pm$0.04 & 0.03$\pm$0.02 \\
    \hline
       PKS 1441+25 & Low & 2.16$\pm$0.13 & 0.018$\pm$0.06 \\ 
        $\bar{\alpha}=2.14\pm0.20$ & High & 1.89$\pm$0.029 & 0.10$\pm$0.02 \\ 
        $\bar{\beta}=0.28\pm0.21$ & Full & 2.08$\pm$0.02 & 0.08$\pm$0.01 \\ 
         & VHE & 1.83$\pm$0.06 & 0.13$\pm$0.04 \\
    \hline
       S3 0218+35 & Low & 2.09$\pm$0.14 &  0.07$\pm$0.07 \\ 
        $\bar{\alpha}=2.28\pm0.21$ & High & 2.32$\pm$0.02 & 0.12$\pm$0.01 \\ 
        $\bar{\beta}=0.20\pm0.16$ & Full & 2.27$\pm$0.01 & 0.056$\pm$0.007 \\ 
         & VHE & 1.62$\pm$0.11 & 0.05$\pm$0.04 \\
    \hline
    \hline
    Source name & State & Index (PL) & \\
    \hline
       B2 1420+32 & Low & 2.42$\pm$0.11 & \\ 
        $\bar{\Gamma}=2.38\pm0.13$ & High & 2.02$\pm$0.01 & \\ 
         & Full & 2.25$\pm$0.01 & \\ 
         & VHE & 2.02$\pm$0.02 & \\
    \hline
       PKS 0346-27 & Low & 2.53$\pm$0.09 & \\ 
        $\bar{\Gamma}=2.22\pm0.09$ & High & 1.99$\pm$0.01 & \\ 
         & Full & 2.097$\pm$0.005 & \\ 
         & VHE & 2.04$\pm$0.04 & \\
    \hline
    \end{tabular}
    \label{tab:Params}
\end{table}


\section{Discussion} \label{sec:Discussion}
We analyzed the Fermi-LAT data of the VHE-detected FSRQs with a focus on the comparative investigation of MeV-GeV spectral and temporal behavior associated with the VHE activity periods to that of non-VHE.
The analysis of the monthly binned light curve (ref. Figure \ref{fig:LC}) shows that all the VHE FSRQs are highly variable, with the ratio of maximum and minimum photon flux being $>100$. Among all sources, PKS 1510-089 exhibits the highest, while TON 0599 exhibits the least variation in the flux amplitude (Table \ref{tab: 4.1}). Since {\it Fermi} continuously and unbiasedly surveys the sky, the number of Bayesian states can be used as a proxy for measuring the source activity, i.e., a greater number of Bayesian bins implies a more active source. Based on this criterion, 3C 279 and PKS 1510-089 are the most active, while PKS 1441+25 is the least active. Also, B2 1420+32 and PKS 0346-27 remain mostly dormant — quiescent/below sensitivity continuously for most of the time ($\sim10$ years).

The photon-flux histograms in Figure \ref{fig:enter-label} are, in general, consistent with both lognormal and normal distributions, though, for a few, one of these seems preferable (p=0.05). Similar conclusions have previously been reported using similar time-bin data \citep[e.g.,][]{2018RAA....18..141S,2023RAA....23k5011W} for a larger sample with a general preference for log-normal. Shorter time-bins incate more complexity \citep[e.g., PKS 2155-304, PKS 1510-089;][]{2010A&A...520A..83H,2016ApJ...822L..13K,2017ApJ...849..138K}.

The VHE emission/activity is generally associated with bright MeV-GeV flux states and has a relatively harder spectrum ($\rm \alpha/\Gamma$: 1.6 -- 2.44) compared to the non-VHE brighter states ($\rm\alpha/\Gamma$: 2.14 -- 2.46). The following spectral trend is apparent:
\begin{itemize}
    \item The MeV-GeV flux is anti-correlated with the spectral index, exhibiting a ``bluer-when-brighter'' trend above a flux limit (Fig. \ref{fig:SIvsFlux} \& \S\ref{sec:SIvsFlux}).
    \item Focusing on a particular source, the high flux spectrum is harder in all except PKS 1510-089 and S3 0218+35 (ref. Table \ref{tab:Params}).
    \item As a whole, the limited sample indicates that the spectral parameter, $\alpha~(\Gamma)$, is systematically lower (a harder spectrum) for high-redshift FSRQs under the assumed spectral model.
\end{itemize}

Within the exposure limit and LAT sensitivity, there appears to be a characteristic difference in the spectral shapes, as seen in Fig. \ref{fig:SEDwithVHE} (bottom panel of each source). The low-state SED is steeper (except for PKS 1510-089, S3 0218+35, and possibly PKS 1441+25), all resembling a power law, implying the high-energy hump peak at $\lesssim300$ MeV.
High-state SEDs, on the other hand, resemble a log parabola with flux in each spectral bin an order of magnitude higher compared to the low state, except for PKS 1501-089.
Examining the subplots in Figure \ref{fig:SEDwithVHE} that shows the ratio of SED data points with respect to maximum SED value (for detected bins only), we found the high-energy peak shows no (e.g., PKS 0736+017 and PKS 1510-089) to a marginal (e.g., 4C+21.35, 3C 279, and S3 0218+35) to a significant (e.g., B2 1420+32, TON 0599, PKS 1441+25, and PKS 0346-27) upshift ($\sim 1-20~GeV$) during high-state compared to the low-state SED. The shift is more prominent in higher redshift FSRQs. 
Such shifts have been reported in some of the VHE FSRQs earlier \citep[e.g., PKS 1441+25, PKS 0736+017, and B2 1420+32;][]{Ahnen2015_pks1441,Abeysekara2015_pks1441,Angioni2019_pks0346,Acciari2021_b2} and also in blazars of other spectral classes \citep[e.g., Mkn 501, 1ES 1215+303, BL Lac;][]{Pian1998,2020ApJ...891..170V,2023MNRAS.521L..53A}.
Overall, the average spectrum for the entire duration is biased towards the higher flux-state spectrum, as expected.

Apart from the SED trend stated above, the VHE-associated MeV-GeV SEDs also show some interesting trends compared to the low- and high-state SEDs. 
\begin{itemize}
    \item The VHE-associated MeV-GeV SED is similar to the low state for PKS 0736+017, PKS 1510-089, 4C+21.35, and 3C 279.
    \item It resembles the high-state SED for B2 1420+32, TON 0599, PKS 1441+25 and PKS 0346-27.
    \item The VHE-associated MeV-GeV SED of S3 0218+35 has a completely different spectral shape, more like an extreme HBL source.
\end{itemize}

These observed spectral changes during the different flux states have strong implications for underlying emission processes and the particle spectrum. Since FSRQs are LSP sources and MeV-GeV emission is almost entirely of leptonic origin via IC, also supported by SED modeling of candidate neutrino blazars \citep[e.g.,][]{Gao}, the observed spectral changes/peak upshift can be primarily due to three reasons: 
\begin{enumerate}[label=(\alph*)]
    \item A change in the underlying particle spectrum.
    \item A change of the dominant seed photon field (indirectly the location of the emission region).
    \item A new additional broadband emission component.
\end{enumerate}
Below, we examine the role of seed photons on the spectral shape, thereby examining its indirect implication to the location of the emission region in the jet.  

\subsection{Emission scenarios}\label{subsec:Emission Scenarios}
For FSRQs, IC-BLR and IC-IR are the most argued scenarios for producing MeV-GeV emissions \citep[e.g., ][]{2018Magic_PKS1510-089}. However, KN disfavors IC-BLR for VHE, and thus only IC-IR (or a strong field with energy $\lesssim$ IR) is relevant. Further, BLR provides strong opacity for $\gamma \gamma$ pair production \citep[e.g.][]{bottcher2016} 
and thus, for VHE production, the emission region is generally argued to be outside the BLR.

The emission may also possibly be from two/multiple emission regions. Since the high-state SED is almost an order of magnitude higher than the low-state SED for each source, for all practical purposes, the high-state emission can be considered effectively from a single emission region. 
For the one-zone scenario, the peak shift reported for the high states is plausible by the IC process through the change in the dominant seed-photon field, indirectly related to the location of the emission region, as demonstrated below.

\subsubsection{Location of emission region and gamma-ray spectrum}\label{subsec:seedPh}
For IC scattering in the Thomson regime, the total power output depends on the energy density of the seed radiation field. 
\begin{equation}
    P(\gamma)~ =~ \frac{4}{3} \sigma_T c \beta^2 \gamma^2 U_R
    \label{eq:Power}
\end{equation}
where $\sigma_T$ is the Thomson scattering cross section, $\beta$ is the speed of the particle in units of the speed of light, $\gamma$ is the particle Lorentz factor, and $U_R$ is the seed photon field energy density. The two most argued cases for the location of the emission region are: within the BLR or outside the BLR.

\textbf{Within BLR:} In this case, both BLR \citep[sub-parsec scale; e.g.][]{2018Natur.563..657G} and IR photons \citep[parsec scale; e.g.][]{2004Natur.429...47J} will be boosted in the blob/emission-region frame (primed quantities), and the respective peak frequencies are given by \citep[e.g.,][]{2012MNRAS.419.1660S}
\begin{equation} \label{eq:nuPin}
        \nu^{P} = \left ( \frac{\delta}{1+z} \right ) \gamma_b'^2 \left ( \Gamma \nu^{*} \right )      
\end{equation}

Thus, for typical SED parameters, the observed IR peak \citep[$T^*$ = 1200 K,][]{Malmarose_2011}\footnote{$h \nu^* = 2.82 k_B T^*$} is 
\begin{equation}
    \nu_{IR}^{P} \approx 2.62\times10^{21} \left ( \frac{\delta}{23.67} \right ) \left ( \frac{\gamma_b'}{340} \right )^2 \left ( \frac{\Gamma}{20.9} \right ) \left ( \frac{\nu_{IR}^{*}}{7.05\times 10^{13}} \right )Hz
\end{equation}

and for BLR (Ly-$\alpha$; $\nu^*=2.47\times10^{15}Hz$), it is
\begin{equation}
\label{eq:vp_blr_in}
       \nu_{BLR}^{P} \approx 9.20\times10^{22} \left ( \frac{\delta}{23.67} \right ) \left ( \frac{\gamma_b'}{340} \right )^2 \left ( \frac{\Gamma}{20.9} \right ) \left ( \frac{\nu_{BLR}^{*}}{2.47\times 10^{15}} \right )Hz
\end{equation}
where $\Gamma$ is the bulk Lorentz factor of the emission region, z is the source redshift, and $\gamma_b'$ is the break Lorentz factor of the 
broken power-law electron distribution used in the SED modeling \citep[e.g.][]{Roy2021}, and $\delta$ is the Doppler factor, $\nu_{IR}^*$ (T$^*$) and $\nu_{BLR}^*$ are respectively the characteristic frequencies of IR and BLR photon fields (assumed thermal) in the AGN frame. Thus, the IC-BLR peak will be at higher frequencies than the IC-IR. 

\begin{figure*}
\centering
\includegraphics[width=.23\linewidth]{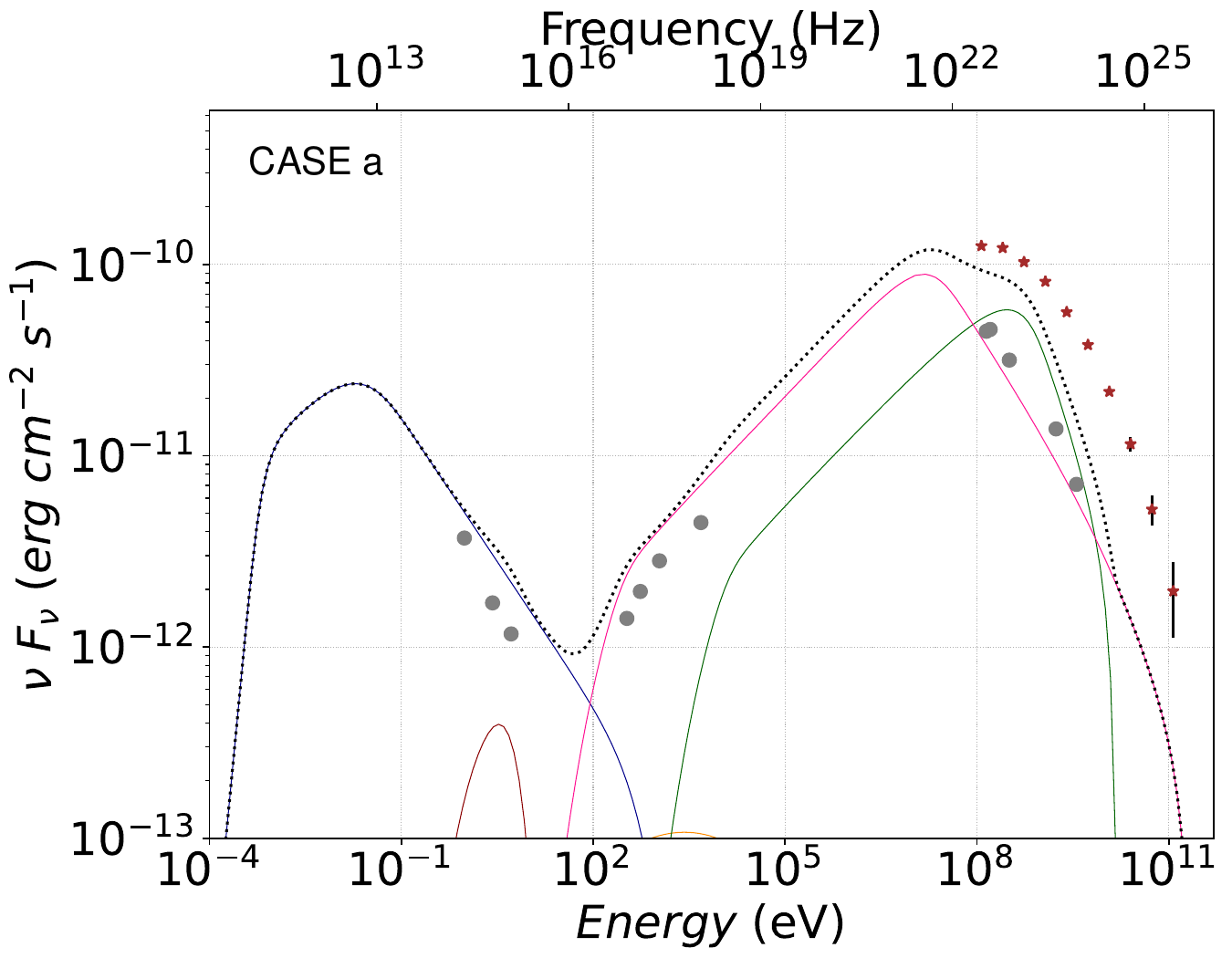}
\includegraphics[width=.23\linewidth]{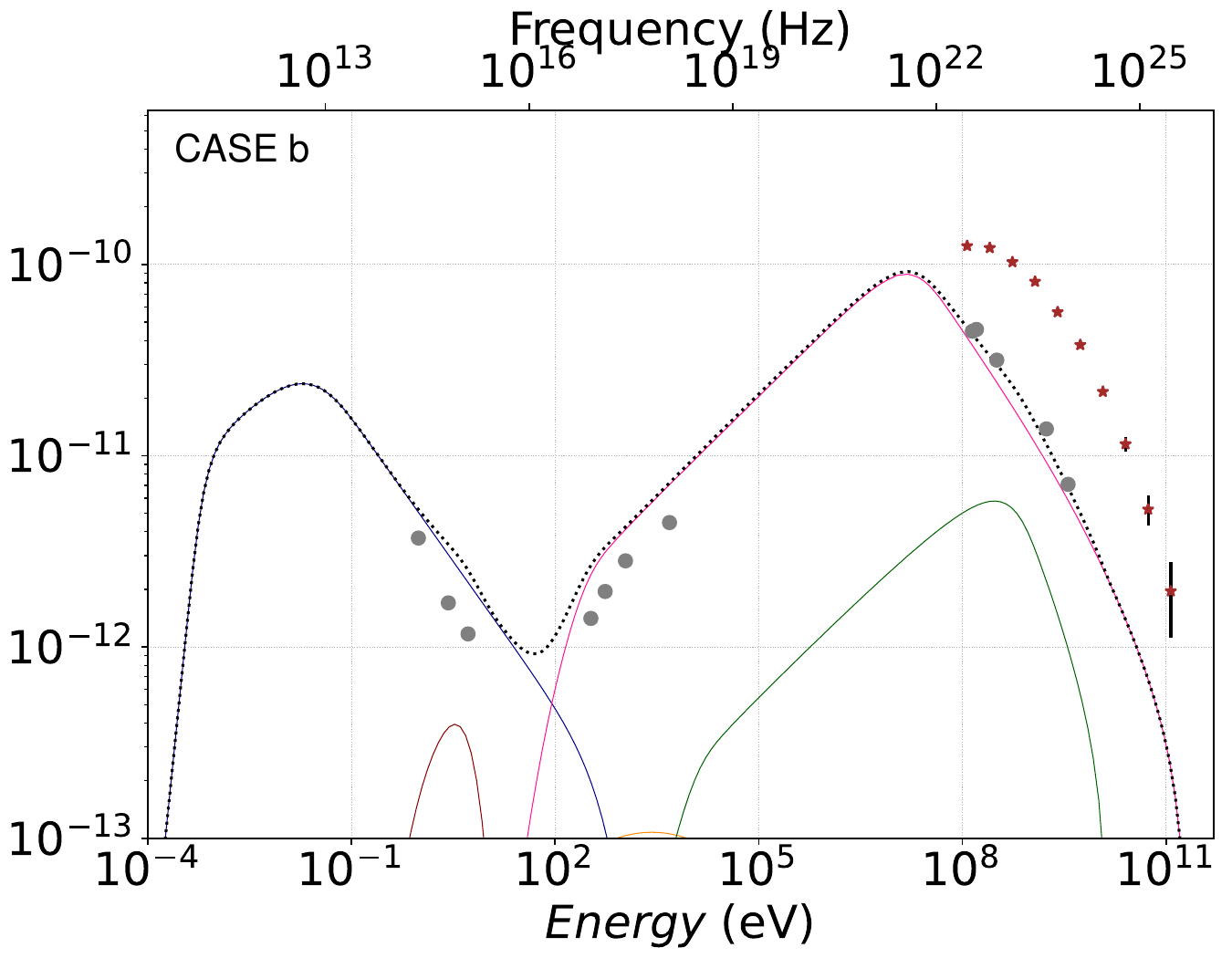}
\includegraphics[width=.23\linewidth]{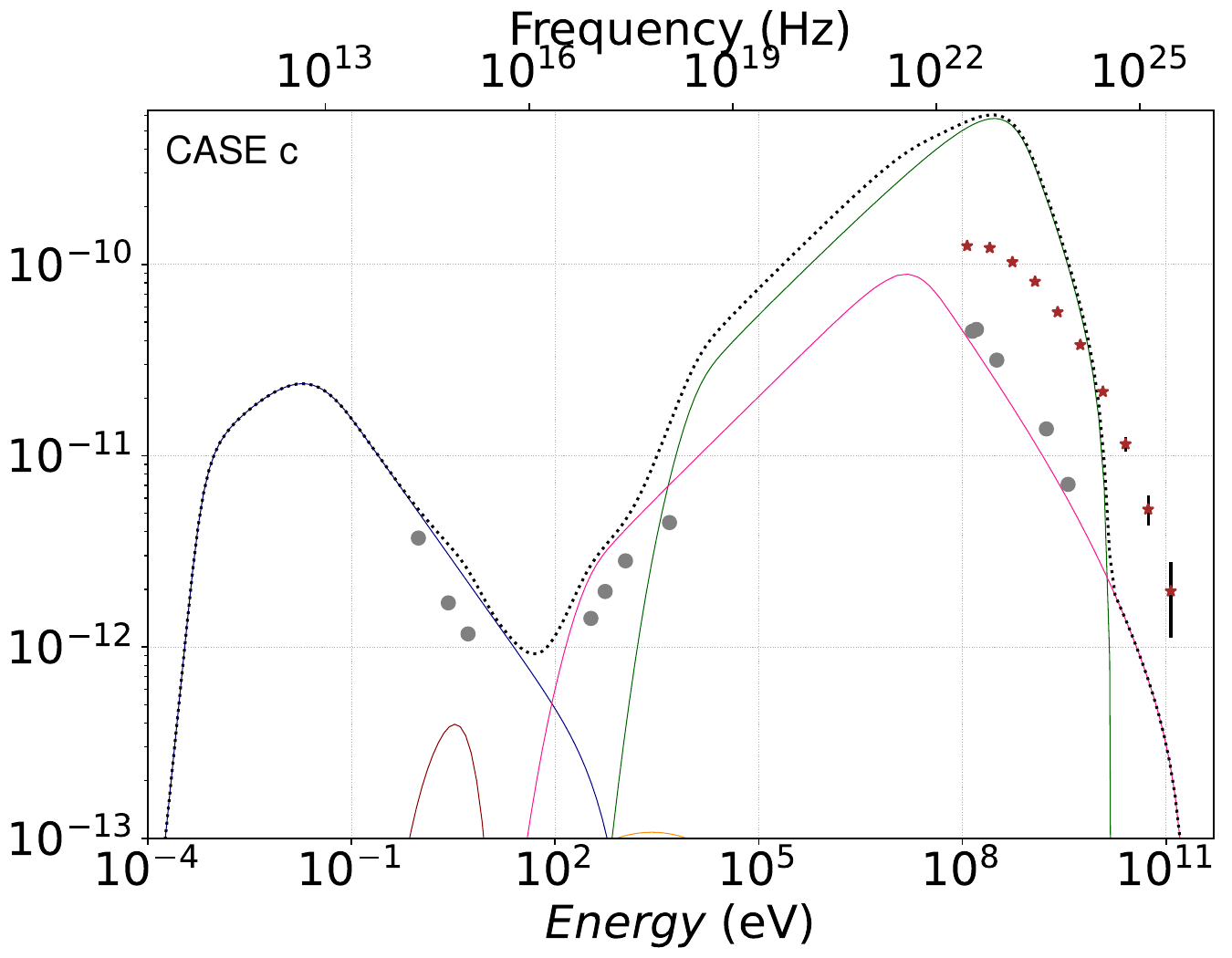}
\includegraphics[width=.23\linewidth]{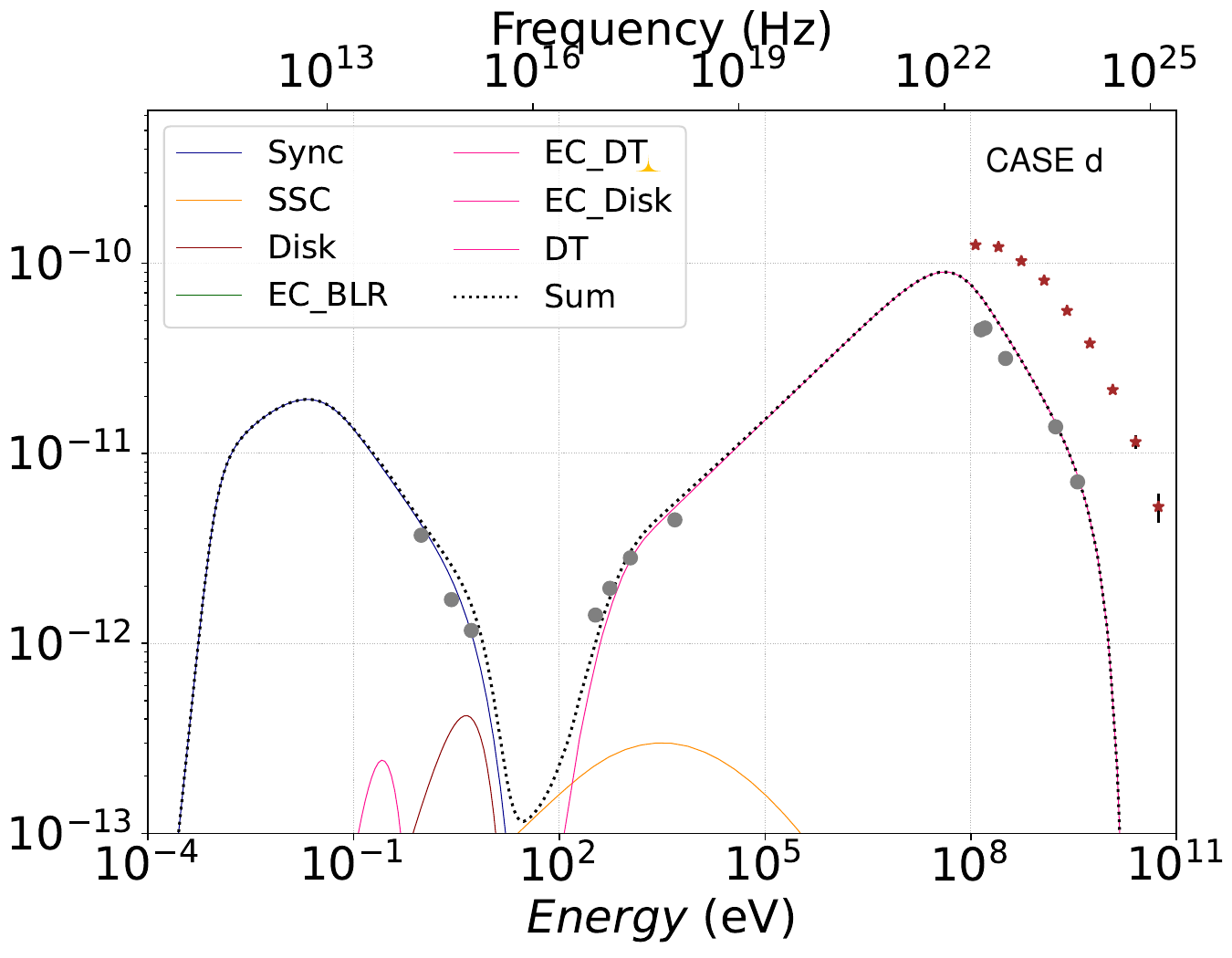}
\caption{Modeled SED for FSRQ 3C 279 demonstrates four different scenarios based on the location of the emission region (ref \ref{subsec:seedPh}). The first three figures correspond to cases (a) to (c) detailed in \S\ref{subsec:seedPh}, where the emission region is within the BLR. The fourth figure depicts the case (d) when the emission region lies outside the BLR. Grey points represent the data points from \citet{Roy2021}, while red data points represent the average spectra for the entire duration. The dark blue, dark green, deep pink, dark red, and dark orange lines indicate synchrotron EC-BLR, EC-IR (EC-DT), disk thermal, and SSC components, respectively. The dashed black line represents the total spectrum. }
\label{fig:sampleSEDpks1510}
\end{figure*}

For a given set of parameters, the radiative power output depends only on the energy density of the seed photons (eq. \ref{eq:Power}). Thus, depending on the relative energy densities of the IR ($U_{IR}$) and BLR ($U_{BLR}$) fields, the MeV-GeV spectrum can exhibit a wide range of SED shapes, like flat, rising, and declining, accompanied by a peak shift. The ratio of their energy density is
\begin{equation}
    \frac{U_{BLR}}{U_{IR}} = \frac{f_{BLR}}{f_{IR}}* \left ( \frac{T_{BLR}}{T_{IR}} \right )^4
\end{equation}
where $f_{IR}$ and $f_{BLR}$ are the overall normalization factors, accounting for the unknown factors like geometry, filling factor, covering fraction, etc. So, we present four cases\footnote{arbitrarily chosen for demonstration only at the moment, but can be used with telescope sensitivity and observed spectral shape to get better constraints}, to demonstrate the effect of varying the energy density of BLR with respect to IR (IR is most relevant for VHE).
This is demonstrated in Figure \ref{fig:sampleSEDpks1510} from case (a) to (d) for the SED of 3C 279 for different $U_R$ using the spectral parameters from \cite{Roy2021}.  \\
(a) $U_{IR}~=~U_{BLR}$, i.e., a similar contribution by both except BLR has an early KN onset and, thus, roughly a flat SED in the sub-GeV to GeV range, followed by a spectrum governed by EC-IR at VHE, if the spectrum continues. This is shown in Fig. \ref{fig:sampleSEDpks1510}(a).  \\
(b) $U_{BLR}~=~0.1*U_{IR}$, improbable in the standard AGN paradigm; can be possible because of the f-factor (clumpy, non-isotropic with the subtended solid angle less than the IR-torus, etc.) and the resulting spectrum is primarily due to EC-IR and peaks at lower energy ($\sim100$ MeV; Fig. \ref{fig:sampleSEDpks1510}(b)). \\
(c) $U_{BLR}~=~10*U_{IR}$, the output is dominated by BLR, leading to a rising spectral shape, peaking at higher energies ($\sim1$ GeV; refer to Fig. \ref{fig:sampleSEDpks1510} (c)). Such SED changes have been seen earlier, e.g., \citet{Hayashida2015}.


(d) \textbf{Outside BLR but within IR torus:} IR photon density will be boosted, while BLR photons will be deboosted. The respective peak frequencies are
\begin{equation}
    \nu_{BLR}^{P,out} = \left ( \frac{\delta}{1+z} \right ) \gamma_b^2 \left ( \frac{\nu_{BLR}^{*}}{\Gamma} \right )
\end{equation}

\begin{equation}
    \nu_{BLR}^{P,out} \approx 2.10\times10^{20} \left ( \frac{\delta}{23.67} \right ) \left ( \frac{\gamma_b}{340} \right )^2 \left ( \frac{20.9}{\Gamma} \right ) \left ( \frac{\nu_{BLR}^{*}}{2.47\times 10^{15}} \right )Hz
\end{equation}
In this case, due to deboosting, BLR will peak at lower energies than IR, and the energy density of BLR will also be lower by a factor of $\Gamma^2$. This case has been shown in figure \ref{fig:sampleSEDpks1510} (d).

In the above demonstrated scenarios, both the IR and BLR fields are in the AGN frame, with the IR field temperature already on the higher side, while the BLR field is Ly-$\alpha$.
Thus, FSRQ's MeV-GeV spectrum in the single emission region scenario is shaped by $\rm U_{IR}$ and $\rm U_{BLR}$, dictated by the location of the emission region. Following the demonstration above, the competitive dominance between the two, even without any upshift in the optical-IR synchrotron peak, can lead to a high-energy peak shift if EC-BLR dominates over EC-IR \citep[e.g.][]{Hayashida2015}, the latter being generally the preferred scenario in LSPs for MeV-GeV gamma-rays (low-state SEDs in Figure \ref{fig:SEDwithVHE}) and VHE emission. However, EC-BLR dominance (emission region within BLR) is ineffective for VHE due to the KN effect, while VHE, which could result from the EC-IR in this case, gets absorbed due to the high opacity of the BLR field through the $\gamma\gamma$ pair creation channel \citep{bottcher2016,Sameer_2024}. Therefore, the VHE is due to EC-IR.

Further, FSRQs' non-thermal optical-IR synchrotron in FSRQs is steeper/softer/declining, and thus, following the trends in Figure \ref{fig:sampleSEDpks1510}, the sub-GeV to GeV spectrum should be similar, i.e., steeper/declining if EC-IR dominates with a peak $\lesssim$100 MeV (cases (b) and (d)); flatter up to $\sim 10-20$ GeV if EC-BLR and EC-IR have a similar contribution (case (a)); or harder with a peak shift to around 10-20 GeV in EC-BLR domination (case (c)). If similar considerations hold regardless of the source redshift, the MeV-GeV peak of high-redshift FSRQs should be systematically at lower energies compared to the low-z counterparts.

From the low, high, and VHE-associated SEDs in Figure \ref{fig:SEDwithVHE}, we have two sets of sources: FSRQs with nil or marginal peak upshift and spectral hardening (Nil: PKS 0736+017, PKS 1510-089; Marginal: 4C+21.35, 3C 279) and FSRQs with substantial peak upshift and hardening (B2 1420+32, TON 0599, PKS 1441+25\footnote{a bit ambiguous due to non-detection at lower energy during the low state.}, S3 0218+35, PKS 0346-27). Noting that IR and BLR seed photons are in the AGN frame, a significant upshift is possible only due to either a significant upshift of the low-energy peak, i.e., the optical-UV synchrotron peak, or a new emission component (\S\ref{subsec:Emission Scenarios}).


For FSRQs with no/marginal peak upshift -- all low-redshift ones, multi-wavelength observations during high-flux and VHE episodes do not indicate significant upshift of the synchrotron peak \citep[e.g.,][]{Ahnen2015_pks1441,Acciari2021_b2,Aharonian_2023,HESS2019_3C279_2015_VHE}. Thus, both the VHE and the high-energy emission are due to EC-IR, as normally argued. However, the spectral hardening seen during VHE as well as in the high states (harder than the VHE spectrum) implies that the VHE emission is primarily a result of a power law extension of the high-energy end of the underlying particle spectrum, i.e., the optical-UV synchrotron spectrum extends to X-ray energies \citep[e.g.,][]{2018Magic_PKS1510-089,Aharonian_2023}.


For FSRQs exhibiting a significant peak upshift (high-redshift ones), the VHE emission can be explained using the normally argued EC-IR scenario if the IR-optical synchrotron peak shows a similar order peak upshift as in the case of PKS 0346-27 \citep{Angioni2019_pks0346} and OP 313 \citep{2024A&A...681A.116P}. For TON 0599, PKS 1441+25, and S3 0218+35, no peak shift in the synchrotron spectrum is reported, and thus we argue that VHE emission is due to an additional HBL/HSP component, which has a Compton dominance (CD) of FSRQ ($CD >> 1$), contrary to the trend inferred from the blazar spectral sequence \citep[$CD \lesssim 1$;][]{blazarClassification,2017MNRAS.469..255G}. For B2 1420+32, it is a combination of both cases: the SEDs reported by \citet{MAGIC2021_B2_2020_VHE} show a significant upshift in the IR/optical spectrum, while the VHE state spectrum shows the need for an additional HBL-like component. The need for an HBL/HSP-like component has already been argued for VHE emission in some of the FSRQs \citep[e.g., PKS 1441+25, S3 0218+35;][]{Ahnen2015_pks1441, Ahnen2016_S3_HBLtpe} and even VHE BL Lacs \citep[e.g., OJ 287:][]{2018MNRAS.479.1672K,2021ApJ...921...18K,2024ApJ...973..134A}. In both cases, the CD of the required new HBL-like component is that of the respective spectral class, indicating that CD is intrinsic.



Though VHE is associated with brighter MeV-GeV flux states, many
(TON 0599, B2 1420+32, PKS 0346-27) have brighter MeV-GeV states and even a harder spectrum compared to the VHE-associated states, but lack VHE association. We argue that the high-state emission in these sources may also be due to an HBL/HSP-like component. For example, for TON 0599, \citet{2025Ton0599_rajPrince} reported a requirement of a thermal accretion-disk component, arguing a signature of disc-jet coupling, to fit the low-energy X-ray spectrum from XMM-Newton, while the MeV-GeV emission is explained via EC-BLR and EC-IR. However, a careful examination of the SEDs shows that the model systematically fails to reproduce the highest MeV-GeV fluxes and also that the authors have not used the XMM-Newton spectra in SED modeling. We argue that the claimed thermal component is actually an HBL/HSP component, manifesting as a thermal-like component because of the limited spectral range. A similar issue can be seen in the SED modeling work of \citet{2024MNRAS.529.1356M}. Thus, even high-flux states SED could be due to an additional HBL-like component, without VHE association, simply because of the particle spectrum. This indicates that a more nuanced look is required during SED modeling.

From the above discussion, it is possible that if the high-energy particle spectrum giving optical-UV synchrotron continued into X-rays \citep[e.g.][]{2025arXiv250405083D,universe11030084}, more FSRQs will be candidate VHE sources via the traditional EC-IR scenario. However, increased redshift implies increased EBL absorption, which requires extraordinary brightening under the EC-IR scenario than what is observed to be detected at VHE. On the contrary, a spectral transition or a new HBL/HSP-spectral component with CD-like FSRQs, though rare, naturally enhances the GeV-VHE
flux to negate EBL absorption. For example, the high redshift of the brightest blazar, 3C 454.3, could be a reason for the lack of VHE detection despite intense flaring behavior \citep[e.g.,][]{2009A&A...498...83A}.

Thus, in short, VHE in FSRQs is primarily driven by the particle spectrum, combined with spectral transition or a new HBL-like component with the CD of FSRQs in high-redshifted ones. Even high-state MeV-GeV without VHE association in a few could be due to an HBL-like component. It should be noted that recent and previous studies have studied VHE FSRQs, e.g. \citet{2025MNRAS.539.2185M}, explored VHE FSRQs with a focus on the statistical properties of flux and spectral index or SED modeling of limited samples, while we focused on comparative SED shapes and their evolution during low, high, and VHE-associated states.

\section{Summary and Conclusion} \label{sec:summary}
VHE-detected FSRQs are relatively few, and therefore, to unravel and understand VHE and high-energy emission, we systematically studied the spectral and temporal properties of the VHE-FSRQs using 14 years of {\it Fermi}-LAT data through monthly binned light curves, exploring statistical properties of flux and spectra and SEDs during different flux states. A summary of our results and inferences is as follows: 



\begin{itemize}
    \item All sources are highly variable, with a maximum-to-minimum flux ratio $>100$ in monthly bins. PKS 1510-089 is the most variable, and TON 0599 is the least variable, based on the Bayesian block.
    
    \item VHE is associated with a brighter MeV-GeV state and, in general, has a relatively harder spectral index. For many, there are brighter MeV-GeV states without VHE. 
    
    \item The logarithmic-flux histograms are consistent with both normal and lognormal distributions, although a few prefer one of the two.
    
    \item Above a flux threshold, the flux anticorrelates with the spectral index, indicating a bluer-when-brighter trend. The lack of this trend at lower flux states is likely due to the detection of fewer photons in a very limited LAT energy range.
    
    \item MeV-GeV SEDs during low states resemble PL, whereas the brighter and VHE-associated SEDs resemble LP. Interestingly, the high-flux state and the VHE SEDs show no (PKS 0736+017, PKS 1510-089), marginal (4C+21.35, 3C 279), or significant peak upshift (B2 1420+32, TON 0599, PKS 1441+25, S3 0218+35, PKS 0346-27, OP 313) compared to the low state. The peak upshift is more prominent in high-redshift sources. 
    

    \item FSRQs with no/marginal peak upshift (all low redshift ones; PKS 0736+017, PKS 1510-089, 4C+21.35, and 3C 279), the low, high, and VHE emission is due to EC-IR with VHE driven by the extension of the high-energy part of the particle spectrum to higher energies. The same is true for FSRQs (PKS 0346-27 and OP 313), where the synchrotron peak has upshifted by a similar order.

    \item For FSRQs with a significant MeV-GeV peak upshift (all high redshift ones; TON 0599, PKS 1441+25, S3 0218+35, PKS 0346-27, OP 313) but no upshift of the synchrotron peak, VHE emission is due to an (extreme) HBL/HSP-like component with a CD of FSRQs (i.e., CD $>>$1), unlike the HBL/HSP CD seen in the blazar sequence (CD$\lesssim$1), powered by an extension of the high-energy end of the particle spectrum to higher energies. For B2 1420+32, it is a combination of both cases.

    \item In high-z FSRQs, even high-flux state SEDs could be from an additional HBL-like component.

\end{itemize}

In a nutshell, VHE is primarily due to a (likely) PL continuation of the high-energy end of the particle spectrum to higher energies supplemented by a spectral transition or a new-jet component with an FSRQ-like CD for high-z FSRQs. The latter may be true even for the high-flux state emission (e.g., TON 0599). For high-z FSRQs, spectral changes naturally boost GeV-VHE flux, negating EBL absorption, which otherwise requires significantly more drastic flux variation than typically seen ($\sim 100$) in the traditional EC-IR scenario.

\appendix

\begin{figure}
\centering
\includegraphics[width=18cm, height=12cm]{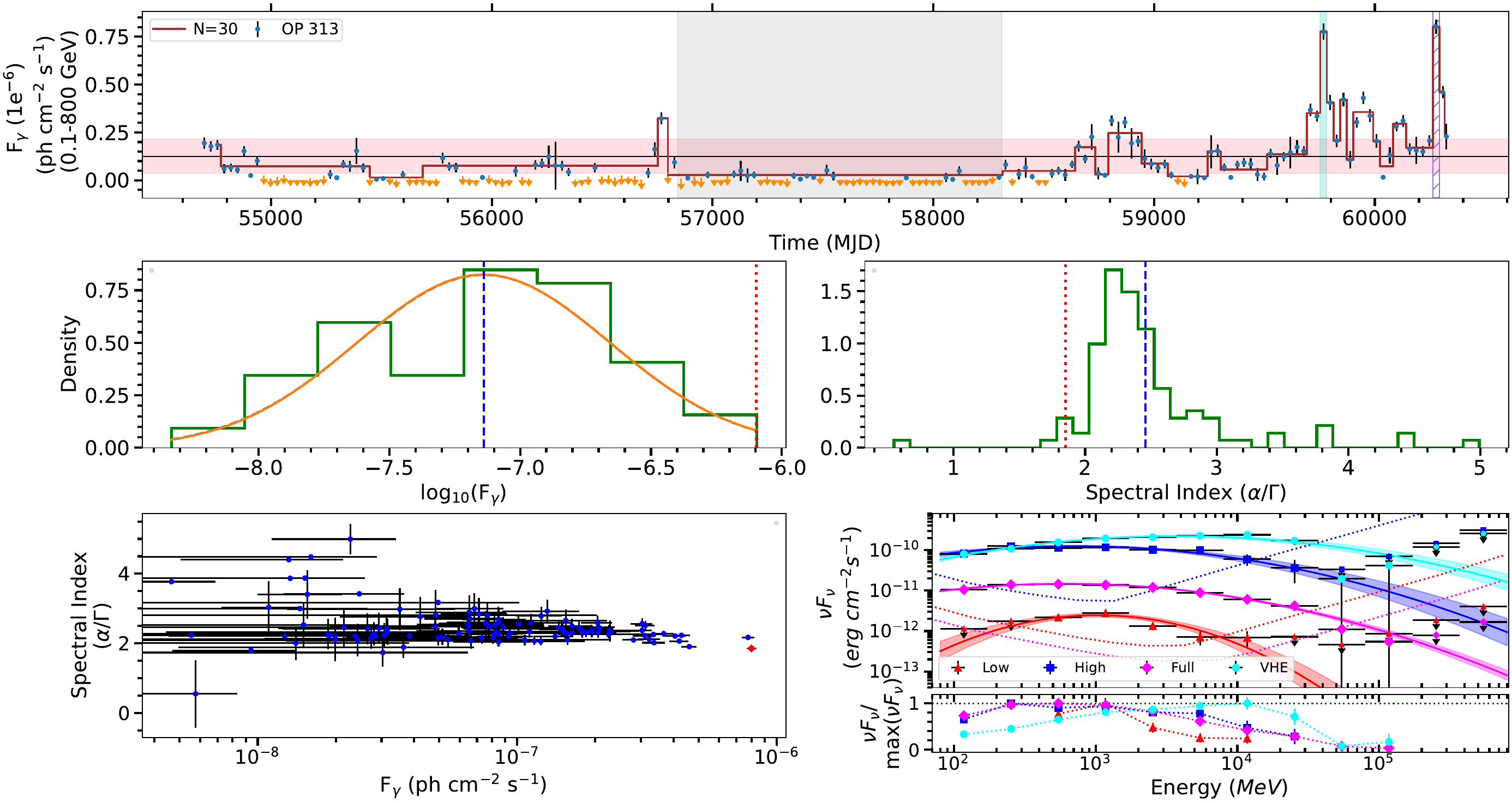}
\caption{Top panel: Monthly binned lightcurve for the 10th VHE FSRQ OP 313. Middle panel: Histogram of logarithmic flux (left) and histogram of spectral index (right). Bottom panel: Spectral index versus photon flux (left) and MeV-GeV SED during low (Low) and high (High) flux states, along with the SED for the entire observation period (Full) and VHE detected durations (VHE; left). The other details are the same as in Figures \ref{fig:LC}, \ref{fig:SIhist}, \ref{fig:SIvsFlux} \ref{fig:SEDwithVHE}.}
\label{fig:SED_OP313}
\end{figure}


\bibliography{references_samplebib}{}

\begin{thebibliography}{}
\expandafter\ifx\csname natexlab\endcsname\relax\def\natexlab#1{#1}\fi
\providecommand{\url}[1]{\href{#1}{#1}}
\providecommand{\dodoi}[1]{doi:~\href{http://doi.org/#1}{\nolinkurl{#1}}}
\providecommand{\doeprint}[1]{\href{http://ascl.net/#1}{\nolinkurl{http://ascl.net/#1}}}
\providecommand{\doarXiv}[1]{\href{https://arxiv.org/abs/#1}{\nolinkurl{https://arxiv.org/abs/#1}}}

\bibitem[{{Abdo} {et~al.}(2010){Abdo}, {Ackermann}, {Agudo}, {Ajello}, {Aller}, {Aller}, {Angelakis}, {Arkharov}, {Axelsson}, {Bach}, {Baldini}, {Ballet}, {Barbiellini}, {Bastieri}, {Baughman}, {Bechtol}, {Bellazzini}, {Benitez}, {Berdyugin}, {Berenji}, {Blandford}, {Bloom}, {Boettcher}, {Bonamente}, {Borgland}, {Bregeon}, {Brez}, {Brigida}, {Bruel}, {Burnett}, {Burrows}, {Buson}, {Caliandro}, {Calzoletti}, {Cameron}, {Capalbi}, {Caraveo}, {Carosati}, {Casandjian}, {Cavazzuti}, {Cecchi}, {{\c{C}}elik}, {Charles}, {Chaty}, {Chekhtman}, {Chen}, {Chiang}, {Chincarini}, {Ciprini}, {Claus}, {Cohen-Tanugi}, {Colafrancesco}, {Cominsky}, {Conrad}, {Costamante}, {Cutini}, {D'ammando}, {Deitrick}, {D'Elia}, {Dermer}, {de Angelis}, {de Palma}, {Digel}, {Donnarumma}, {Silva}, {Drell}, {Dubois}, {Dultzin}, {Dumora}, {Falcone}, {Farnier}, {Favuzzi}, {Fegan}, {Focke}, {Forn{\'e}}, {Fortin}, {Frailis}, {Fuhrmann}, {Fukazawa}, {Funk}, {Fusco}, {G{\'o}mez}, {Gargano}, {Gasparrini}, {Gehrels}, {Germani}, {Giebels}, {Giglietto},
  {Giommi}, {Giordano}, {Giuliani}, {Glanzman}, {Godfrey}, {Grenier}, {Gronwall}, {Grove}, {Guillemot}, {Guiriec}, {Gurwell}, {Hadasch}, {Hanabata}, {Harding}, {Hayashida}, {Hays}, {Healey}, {Heidt}, {Hiriart}, {Horan}, {Hoversten}, {Hughes}, {Itoh}, {Jackson}, {J{\'o}hannesson}, {Johnson}, {Johnson}, {Jorstad}, {Kadler}, {Kamae}, {Katagiri}, {Kataoka}, {Kawai}, {Kennea}, {Kerr}, {Kimeridze}, {Kn{\"o}dlseder}, {Kocian}, {Kopatskaya}, {Koptelova}, {Konstantinova}, {Kovalev}, {Kovalev}, {Kurtanidze}, {Kuss}, {Lande}, {Larionov}, {Latronico}, {Leto}, {Lindfors}, {Longo}, {Loparco}, {Lott}, {Lovellette}, {Lubrano}, {Madejski}, {Makeev}, {Marchegiani}, {Marscher}, {Marshall}, {Max-Moerbeck}, {Mazziotta}, {McConville}, {McEnery}, {Meurer}, {Michelson}, {Mitthumsiri}, {Mizuno}, {Moiseev}, {Monte}, {Monzani}, {Morselli}, {Moskalenko}, {Murgia}, {Nestoras}, {Nilsson}, {Nizhelsky}, {Nolan}, {Norris}, {Nuss}, {Ohsugi}, {Ojha}, {Omodei}, {Orlando}, {Ormes}, {Osborne}, {Ozaki}, {Pacciani}, {Padovani}, {Pagani}, {Page},
  {Paneque}, {Panetta}, {Parent}, {Pasanen}, {Pavlidou}, {Pelassa}, {Pepe}, {Perri}, {Pesce-Rollins}, {Piranomonte}, {Piron}, {Pittori}, {Porter}, {Puccetti}, {Rahoui}, {Rain{\`o}}, {Raiteri}, {Rando}, {Razzano}, {Reimer}, {Reimer}, {Reposeur}, {Richards}, {Ritz}, {Rochester}, {Rodriguez}, {Romani}, {Ros}, {Roth}, {Roustazadeh}, {Ryde}, {Sadrozinski}, {Sadun}, {Sanchez}, {Sander}, {Saz Parkinson}, {Scargle}, {Sellerholm}, {Sgr{\`o}}, {Shaw}, {Sigua}, {Siskind}, {Smith}, {Smith}, {Spandre}, {Spinelli}, {Starck}, {Stevenson}, {Stratta}, {Strickman}, {Suson}, {Tajima}, {Takahashi}, {Takahashi}, {Takalo}, {Tanaka}, {Thayer}, {Thayer}, {Thompson}, {Tibaldo}, {Torres}, {Tosti}, {Tramacere}, {Uchiyama}, {Usher}, {Vasileiou}, {Verrecchia}, {Vilchez}, {Villata}, {Vitale}, {Waite}, {Wang}, {Winer}, {Wood}, {Ylinen}, {Zensus}, {Zhekanis}, \& {Ziegler}}]{SEDdoubleHump}
{Abdo}, A.~A., {Ackermann}, M., {Agudo}, I., {et~al.} 2010, \apj, 716, 30, \dodoi{10.1088/0004-637X/716/1/30}

\bibitem[{Abdollahi {et~al.}(2022)Abdollahi, Acero, Baldini, Ballet, Bastieri, Bellazzini, Berenji, Berretta, Bissaldi, Blandford, Bloom, Bonino, Brill, Britto, Bruel, Burnett, Buson, Cameron, Caputo, Caraveo, Castro, Chaty, Cheung, Chiaro, Cibrario, Ciprini, Coronado-Blázquez, Crnogorcevic, Cutini, D’Ammando, De~Gaetano, Digel, Di~Lalla, Dirirsa, Di~Venere, Domínguez, Fallah~Ramazani, Fegan, Ferrara, Fiori, Fleischhack, Franckowiak, Fukazawa, Funk, Fusco, Galanti, Gammaldi, Gargano, Garrappa, Gasparrini, Giacchino, Giglietto, Giordano, Giroletti, Glanzman, Green, Grenier, Grondin, Guillemot, Guiriec, Gustafsson, Harding, Hays, Hewitt, Horan, Hou, Jóhannesson, Karwin, Kayanoki, Kerr, Kuss, Landriu, Larsson, Latronico, Lemoine-Goumard, Li, Liodakis, Longo, Loparco, Lott, Lubrano, Maldera, Malyshev, Manfreda, Martí-Devesa, Mazziotta, Mereu, Meyer, Michelson, Mirabal, Mitthumsiri, Mizuno, Moiseev, Monzani, Morselli, Moskalenko, Negro, Nuss, Omodei, Orienti, Orlando, Paneque, Pei, Perkins, Persic,
  Pesce-Rollins, Petrosian, Pillera, Poon, Porter, Principe, Rainò, Rando, Rani, Razzano, Razzaque, Reimer, Reimer, Reposeur, Sánchez-Conde, Saz~Parkinson, Scotton, Serini, Sgrò, Siskind, Smith, Spandre, Spinelli, Sueoka, Suson, Tajima, Tak, Thayer, Thompson, Torres, Troja, Valverde, Wood, \& Zaharijas}]{Abdollahi2022_4fglCatB}
Abdollahi, S., Acero, F., Baldini, L., {et~al.} 2022, The Astrophysical Journal Supplement Series, 260, 53, \dodoi{10.3847/1538-4365/ac6751}

\bibitem[{{Abeysekara} {et~al.}(2015){Abeysekara}, {Archambault}, {Archer}, {Aune}, {Barnacka}, {Benbow}, {Bird}, {Biteau}, {Buckley}, {Bugaev}, {Cardenzana}, {Cerruti}, {Chen}, {Christiansen}, {Ciupik}, {Connolly}, {Coppi}, {Cui}, {Dickinson}, {Dumm}, {Eisch}, {Errando}, {Falcone}, {Feng}, {Finley}, {Fleischhack}, {Flinders}, {Fortin}, {Fortson}, {Furniss}, {Gillanders}, {Griffin}, {Grube}, {Gyuk}, {H{\"u}tten}, {H{\r{a}}kansson}, {Hanna}, {Holder}, {Humensky}, {Johnson}, {Kaaret}, {Kar}, {Kelley-Hoskins}, {Khassen}, {Kieda}, {Krause}, {Krennrich}, {Kumar}, {Lang}, {Maier}, {McArthur}, {McCann}, {Meagher}, {Moriarty}, {Mukherjee}, {Nieto}, {O'Faol{\'a}in de Bhr{\'o}ithe}, {Ong}, {Otte}, {Park}, {Perkins}, {Petrashyk}, {Pohl}, {Popkow}, {Pueschel}, {Quinn}, {Ragan}, {Ratliff}, {Reynolds}, {Richards}, {Roache}, {Rousselle}, {Santander}, {Sembroski}, {Shahinyan}, {Smith}, {Staszak}, {Telezhinsky}, {Todd}, {Tucci}, {Tyler}, {Vassiliev}, {Vincent}, {Wakely}, {Weiner}, {Weinstein}, {Wilhelm}, {Williams}, {Zitzer},
  {VERITAS}, {Smith}, {SPOL}, {Holoien}, {Prieto}, {Kochanek}, {Stanek}, {Shappee}, {ASAS-SN}, {Hovatta}, {Max-Moerbeck}, {Pearson}, {Reeves}, {Richards}, {Readhead}, {OVRO}, {Madejski}, {NuSTAR}, {Djorgovski}, {Drake}, {Graham}, {Mahabal}, \& {CRTS}}]{Abeysekara2015_pks1441}
{Abeysekara}, A.~U., {Archambault}, S., {Archer}, A., {et~al.} 2015, \apjl, 815, L22, \dodoi{10.1088/2041-8205/815/2/L22}

\bibitem[{Acciari {et~al.}(2021)Acciari, Ansoldi, Antonelli, Arbet~Engels, Artero, Asano, Baack, Babić, Baquero, Barres~de Almeida, Barrio, Becerra~González, Bednarek, Bellizzi, Bernardini, Bernardos, Berti, Besenrieder, Bhattacharyya, Bigongiari, Biland, Blanch, Bonnoli, Bošnjak, Busetto, Carosi, Ceribella, Cerruti, Chai, Chilingarian, Cikota, Colak, Colombo, Contreras, Cortina, Covino, D’Amico, D’Elia, Da~Vela, Dazzi, De~Angelis, De~Lotto, Delfino, Delgado, Delgado~Mendez, Depaoli, Di~Pierro, Di~Venere, Do~Souto~Espiñeira, Dominis~Prester, Donini, Dorner, Doro, Elsaesser, Fallah~Ramazani, Fattorini, Ferrara, Foffano, Fonseca, Font, Fruck, Fukami, García~López, Garczarczyk, Gasparyan, Gaug, Giglietto, Giordano, Gliwny, Godinović, Green, Green, Hadasch, Hahn, Heckmann, Herrera, Hoang, Hrupec, Hütten, Inada, Inoue, Ishio, Iwamura, Jormanainen, Jouvin, Kajiwara, Karjalainen, Kerszberg, Kobayashi, Kubo, Kushida, Lamastra, Lelas, Leone, Lindfors, Lombardi, Longo, López-Coto, López-Moya,
  López-Oramas, Loporchio, Machado~de Oliveira~Fraga, Maggio, Majumdar, Makariev, Mallamaci, Maneva, Manganaro, Mannheim, Maraschi, Mariotti, Martínez, Mazin, Mender, Mićanović, Miceli, Miener, Minev, Miranda, Mirzoyan, Molina, Moralejo, Morcuende, Moreno, Moretti, Neustroev, Nigro, Nilsson, Ninci, Nishijima, Noda, Nozaki, Ohtani, Oka, Otero-Santos, Paiano, Palatiello, Paneque, Paoletti, Paredes, Pavletić, Peñil, Perennes, Persic, Prada~Moroni, Prandini, Priyadarshi, Puljak, Rhode, Ribó, Rico, Righi, Rugliancich, Saha, Sahakyan, Saito, Sakurai, Satalecka, Saturni, Schleicher, Schmidt, Schweizer, Sitarek, Šnidarić, Sobczynska, Spolon, Stamerra, Strom, Strzys, Suda, Surić, Takahashi, Tavecchio, Temnikov, Terzić, Teshima, Torres-Albà, Tosti, Truzzi, Tutone, van Scherpenberg, Vanzo, Vazquez~Acosta, Ventura, Verguilov, Vigorito, Vitale, Vovk, Will, Zarić, Angioni, D’Ammando, Ciprini, Cheung, Orienti, Pacciani, Prajapati, Kumar, Ganesh, Minev, Kurtenkov, Marchini, Carrasco, Escobedo, Porras,
  Recillas, Lähteenmäki, Tornikoski, Berton, Tammi, Vera, Jorstad, Marscher, Weaver, Hart, Hallum, Larionov, Borman, Grishina, Kopatskaya, Larionova, Nikiforova, Morozova, Savchenko, Troitskaya, Troitsky, Vasilyev, Hodges, Hovatta, Kiehlmann, Max-Moerbeck, Readhead, Reeves, \& Pearson}]{Acciari2021_b2}
Acciari, V.~A., Ansoldi, S., Antonelli, L.~A., {et~al.} 2021, Astronomy \&amp; Astrophysics, 647, A163, \dodoi{10.1051/0004-6361/202039687}

\bibitem[{Acharyya {et~al.}(2023)Acharyya, Adams, Archer, Bangale, Bartkoske, Batista, Benbow, Brill, Buckley, Christiansen, Chromey, Errando, Falcone, Feng, Foote, Fortson, Furniss, Gallagher, Hanlon, Hanna, Hervet, Hinrichs, Hoang, Holder, Humensky, Jin, Kaaret, Kertzman, Kherlakian, Kieda, Kleiner, Korzoun, Kumar, Lang, Lundy, Maier, McGrath, Millard, Millis, Mooney, Moriarty, Mukherjee, O’Brien, Ong, Pohl, Pueschel, Quinn, Ragan, Reynolds, Ribeiro, Roache, Sadeh, Sadun, Saha, Santander, Sembroski, Shang, Splettstoesser, Talluri, Tucci, Vassiliev, Weinstein, Williams, Wong, Woo, Collaboration, Aharonian, Aschersleben, Backes, Martins, Batzofin, Becherini, Berge, Bernlöhr, Bi, Böttcher, Boisson, Bolmont, de~Bony~de Lavergne, Borowska, Bouyahiaoui, Bradascio, Breuhaus, Brose, Brun, Bruno, Bulik, Burger-Scheidlin, Caroff, Casanova, Cecil, Celic, Cerruti, Chand, Chandra, Chen, Chibueze, Chibueze, Cotter, Dai, Mbarubucyeye, Djannati-Ataï, Dmytriiev, Doroshenko, Einecke, Ernenwein, de~Clairfontaine,
  Filipovic, Fontaine, Füßling, Funk, Gabici, Ghafourizadeh, Giavitto, Glawion, Glicenstein, Goswami, Grolleron, Haerer, Hinton, Holch, Holler, Horns, Jamrozy, Jankowsky, Joshi, Jung-Richardt, Kasai, Katarzyński, Khatoon, Khélifi, Klepser, Kluźniak, Kosack, Kostunin, Lang, Stum, Lemière, Lenain, Leuschner, Lohse, Luashvili, Lypova, Mackey, Malyshev, Marandon, Marchegiani, Marcowith, Martí-Devesa, Marx, Mitchell, Moderski, Mohrmann, Montanari, Moulin, Murach, Nakashima, Niemiec, Noel, O’Brien, Olivera-Nieto, de~Ona~Wilhelmi, Ostrowski, Panny, Panter, Peron, Prokhorov, Pühlhofer, Punch, Quirrenbach, Reichherzer, Reimer, Reimer, Ren, Renaud, Rieger, Rudak, Ruiz-Velasco, Sahakian, Santangelo, Sasaki, Schäfer, Schüssler, Schutte, Schwanke, Shapopi, Specovius, Spencer, Stawarz, Steenkamp, Steinmassl, Sushch, Suzuki, Takahashi, Tanaka, Terrier, van Eldik, Vecchi, Veh, Venter, Vink, White, Wierzcholska, Wong, Zacharias, Zargaryan, Zdziarski, Zech, Zouari, Żywucka, Collaboration, \& Mori}]{Acharyya_2023}
Acharyya, A., Adams, C.~B., Archer, A., {et~al.} 2023, The Astrophysical Journal, 954, 70, \dodoi{10.3847/1538-4357/ace327}

\bibitem[{{Acharyya} {et~al.}(2024){Acharyya}, {Adams}, {Archer}, {Bangale}, {Bartkoske}, {Batista}, {Benbow}, {Brill}, {Caldwell}, {Carini}, {Christiansen}, {Chromey}, {Errando}, {Falcone}, {Feng}, {Finley}, {Foote}, {Fortson}, {Furniss}, {Gallagher}, {Hanlon}, {Hanna}, {Hervet}, {Hinrichs}, {Hoang}, {Holder}, {Humensky}, {Jin}, {Johnson}, {Kaaret}, {Kertzman}, {Kherlakian}, {Kieda}, {Kleiner}, {Korzoun}, {Krennrich}, {Kumar}, {Lang}, {Lundy}, {Maier}, {McGrath}, {Millard}, {Millis}, {Mooney}, {Moriarty}, {Mukherjee}, {O'Brien}, {Ong}, {Pohl}, {Pueschel}, {Quinn}, {Rabinowitz}, {Ragan}, {Reynolds}, {Ribeiro}, {Roache}, {Ryan}, {Sadeh}, {Sadun}, {Saha}, {Santander}, {Sembroski}, {Shahinyan}, {Shang}, {Splettstoesser}, {Tak}, {Talluri}, {Tucci}, {Williams}, {Wong}, {VERITAS Collaboration}, {Jorstad}, {Lico}, {Lusen}, \& {Marscher}}]{2024ApJ...973..134A}
{Acharyya}, A., {Adams}, C.~B., {Archer}, A., {et~al.} 2024, \apj, 973, 134, \dodoi{10.3847/1538-4357/ad64d0}

\bibitem[{{Agarwal} {et~al.}(2023){Agarwal}, {Banerjee}, {Shukla}, {Roy}, {Acharya}, {Vaidya}, {Chitnis}, {Wagner}, {Mannheim}, \& {Branchesi}}]{2023MNRAS.521L..53A}
{Agarwal}, S., {Banerjee}, B., {Shukla}, A., {et~al.} 2023, \mnras, 521, L53, \dodoi{10.1093/mnrasl/slad023}

\bibitem[{Aharonian {et~al.}(2023)Aharonian, Benkhali, Aschersleben, Ashkar, Backes, Martins, Barnard, Batzofin, Becherini, Berge, Bernlöhr, Bi, de~Bony~de Lavergne, Böttcher, Boisson, Bolmont, Borowska, Bouyahiaoui, Bradascio, Breuhaus, Brose, Brown, Brun, Bruno, Bulik, Burger-Scheidlin, Caroff, Casanova, Cecil, Celic, Cerruti, Chand, Chandra, Chen, Chibueze, Chibueze, Cotter, Damascene~Mbarubucyeye, Davids, Djannati-Ataï, Dmytriiev, Doroshenko, Egberts, Einecke, Ernenwein, Fegan, Fontaine, Füßling, Funk, Gabici, Ghafourizadeh, Giavitto, Glawion, Glicenstein, Goswami, Grolleron, Haerer, Hofmann, Holch, Holler, Horns, Jamrozy, Jankowsky, Joshi, Jung-Richardt, Kasai, Katarzyński, Khatoon, Khélifi, Kluźniak, Komin, Kosack, Kostunin, Lang, Le~Stum, Leitl, Lemière, Lenain, Leuschner, Luashvili, Mackey, Marandon, Marchegiani, Martí-Devesa, Marx, Mehta, Meyer, Mitchell, Moderski, Mohrmann, Montanari, Moulin, de~Naurois, Niemiec, Priyana~Noel, O’Brien, Ohm, Olivera-Nieto, de~Ona~Wilhelmi, Ostrowski,
  Panny, Panter, Peron, Prokhorov, Pühlhofer, Punch, Quirrenbach, Reichherzer, Reimer, Reimer, Ren, Rieger, Rowell, Rudak, Ricarte, Ruiz-Velasco, Sahakian, Salzmann, Sanchez, Santangelo, Sasaki, Schüssler, Schutte, Schwanke, Shapopi, Sol, Specovius, Spencer, Stawarz, Steenkamp, Steinmassl, Steppa, Sushch, Suzuki, Takahashi, Tanaka, Terrier, Tsuji, van Eldik, van Soelen, Vecchi, Veh, Vink, Wach, Wagner, Wierzcholska, Zacharias, Zargaryan, Zdziarski, Zech, Zouari, Żywucka, Buckley, Cooper, \& Groenewald}]{Aharonian_2023}
Aharonian, F., Benkhali, F.~A., Aschersleben, J., {et~al.} 2023, The Astrophysical Journal Letters, 952, L38, \dodoi{10.3847/2041-8213/ace3c0}

\bibitem[{{Ahnen} {et~al.}(2015){Ahnen}, {Ansoldi}, {Antonelli}, {Antoranz}, {Babic}, {Banerjee}, {Bangale}, {Barres de Almeida}, {Barrio}, {Bednarek}, {Bernardini}, {Biasuzzi}, {Biland}, {Blanch}, {Bonnefoy}, {Bonnoli}, {Borracci}, {Bretz}, {Carmona}, {Carosi}, {Chatterjee}, {Clavero}, {Colin}, {Colombo}, {Contreras}, {Cortina}, {Covino}, {Da Vela}, {Dazzi}, {De Angelis}, {De Lotto}, {de O{\~n}a Wilhelmi}, {Delgado Mendez}, {Di Pierro}, {Dominis Prester}, {Dorner}, {Doro}, {Einecke}, {Eisenacher Glawion}, {Elsaesser}, {Fern{\'a}ndez-Barral}, {Fidalgo}, {Fonseca}, {Font}, {Frantzen}, {Fruck}, {Galindo}, {Garc{\'\i}a L{\'o}pez}, {Garczarczyk}, {Garrido Terrats}, {Gaug}, {Giammaria}, {Godinovi{\'c}}, {Gonz{\'a}lez Mu{\~n}oz}, {Guberman}, {Hahn}, {Hanabata}, {Hayashida}, {Herrera}, {Hose}, {Hrupec}, {Hughes}, {Idec}, {Kodani}, {Konno}, {Kubo}, {Kushida}, {La Barbera}, {Lelas}, {Lindfors}, {Lombardi}, {L{\'o}pez}, {L{\'o}pez-Coto}, {L{\'o}pez-Oramas}, {Lorenz}, {Majumdar}, {Makariev}, {Mallot}, {Maneva},
  {Manganaro}, {Mannheim}, {Maraschi}, {Marcote}, {Mariotti}, {Mart{\'\i}nez}, {Mazin}, {Menzel}, {Miranda}, {Mirzoyan}, {Moralejo}, {Moretti}, {Nakajima}, {Neustroev}, {Niedzwiecki}, {Nievas Rosillo}, {Nilsson}, {Nishijima}, {Noda}, {Orito}, {Overkemping}, {Paiano}, {Palacio}, {Palatiello}, {Paneque}, {Paoletti}, {Paredes}, {Paredes-Fortuny}, {Persic}, {Poutanen}, {Prada Moroni}, {Prandini}, {Puljak}, {Rhode}, {Rib{\'o}}, {Rico}, {Rodriguez Garcia}, {Saito}, {Satalecka}, {Schultz}, {Schweizer}, {Shore}, {Sillanp{\"a}{\"a}}, {Sitarek}, {Snidaric}, {Sobczynska}, {Stamerra}, {Steinbring}, {Strzys}, {Takalo}, {Takami}, {Tavecchio}, {Temnikov}, {Terzi{\'c}}, {Tescaro}, {Teshima}, {Thaele}, {Torres}, {Toyama}, {Treves}, {Verguilov}, {Vovk}, {Ward}, {Will}, {Wu}, {Zanin}, {MAGIC Collaboration}, {Ajello}, {Baldini}, {Barbiellini}, {Bastieri}, {Becerra Gonz{\'a}lez}, {Bellazzini}, {Bissaldi}, {Blandford}, {Bonino}, {Bregeon}, {Bruel}, {Buson}, {Caliandro}, {Cameron}, {Caragiulo}, {Caraveo}, {Cavazzuti}, {Chiang},
  {Chiaro}, {Ciprini}, {D'Ammando}, {de Palma}, {Desiante}, {Di Venere}, {Dom{\'\i}nguez}, {Fusco}, {Gargano}, {Gasparrini}, {Giglietto}, {Giordano}, {Giroletti}, {Grenier}, {Guiriec}, {Hays}, {Hewitt}, {Jogler}, {Kuss}, {Larsson}, {Li}, {Li}, {Longo}, {Loparco}, {Lovellette}, {Lubrano}, {Maldera}, {Mayer}, {Mazziotta}, {McEnery}, {Mirabal}, {Mizuno}, {Monzani}, {Morselli}, {Moskalenko}, {Nuss}, {Ojha}, {Ohsugi}, {Omodei}, {Orlando}, {Perkins}, {Pesce-Rollins}, {Piron}, {Pivato}, {Porter}, {Raino}, {Rando}, {Razzano}, {Reimer}, {Reimer}, {Sgro}, {Siskind}, {Spada}, {Spandre}, {Spinelli}, {Tajima}, {Takahashi}, {Thayer}, {Thompson}, {Troja}, {Wood}, {Fermi-LAT Collaboration}, {Balokovic}, {Berdyugin}, {Carraminana}, {Carrasco}, {Chavushyan}, {Fallah Ramazani}, {Feige}, {Haarto}, {Haeusner}, {Hovatta}, {Kania}, {Klamt}, {L{\"a}hteenm{\"a}ki}, {Leon-Tavares}, {Lorey}, {Pacciani}, {Porras}, {Recillas}, {Reinthal}, {Tornikoski}, {Wolfert}, \& {Zottmann}}]{Ahnen2015_pks1441}
{Ahnen}, M.~L., {Ansoldi}, S., {Antonelli}, L.~A., {et~al.} 2015, \apjl, 815, L23, \dodoi{10.1088/2041-8205/815/2/L23}

\bibitem[{{Ahnen} {et~al.}(2016){Ahnen}, {Ansoldi}, {Antonelli}, {Antoranz}, {Arcaro}, {Babic}, {Banerjee}, {Bangale}, {Barres de Almeida}, {Barrio}, {Becerra Gonz{\'a}lez}, {Bednarek}, {Bernardini}, {Berti}, {Biasuzzi}, {Biland}, {Blanch}, {Bonnefoy}, {Bonnoli}, {Borracci}, {Bretz}, {Buson}, {Carosi}, {Chatterjee}, {Clavero}, {Colin}, {Colombo}, {Contreras}, {Cortina}, {Covino}, {Da Vela}, {Dazzi}, {De Angelis}, {De Lotto}, {de O{\~n}a Wilhelmi}, {Di Pierro}, {Doert}, {Dom{\'\i}nguez}, {Dominis Prester}, {Dorner}, {Doro}, {Einecke}, {Eisenacher Glawion}, {Elsaesser}, {Engelkemeier}, {Fallah Ramazani}, {Fern{\'a}ndez-Barral}, {Fidalgo}, {Fonseca}, {Font}, {Frantzen}, {Fruck}, {Galindo}, {Garc{\'\i}a L{\'o}pez}, {Garczarczyk}, {Garrido Terrats}, {Gaug}, {Giammaria}, {Godinovi{\'c}}, {Gora}, {Guberman}, {Hadasch}, {Hahn}, {Hayashida}, {Herrera}, {Hose}, {Hrupec}, {Hughes}, {Idec}, {Kodani}, {Konno}, {Kubo}, {Kushida}, {La Barbera}, {Lelas}, {Lindfors}, {Lombardi}, {Longo}, {L{\'o}pez}, {L{\'o}pez-Coto},
  {Majumdar}, {Makariev}, {Mallot}, {Maneva}, {Manganaro}, {Mannheim}, {Maraschi}, {Marcote}, {Mariotti}, {Mart{\'\i}nez}, {Mazin}, {Menzel}, {Miranda}, {Mirzoyan}, {Moralejo}, {Moretti}, {Nakajima}, {Neustroev}, {Niedzwiecki}, {Nievas Rosillo}, {Nilsson}, {Nishijima}, {Noda}, {Nogu{\'e}s}, {Paiano}, {Palacio}, {Palatiello}, {Paneque}, {Paoletti}, {Paredes}, {Paredes-Fortuny}, {Pedaletti}, {Peresano}, {Perri}, {Persic}, {Poutanen}, {Prada Moroni}, {Prandini}, {Puljak}, {Garcia}, {Reichardt}, {Rhode}, {Rib{\'o}}, {Rico}, {Saito}, {Satalecka}, {Schroeder}, {Schweizer}, {Shore}, {Sillanp{\"a}{\"a}}, {Sitarek}, {Snidaric}, {Sobczynska}, {Stamerra}, {Strzys}, {Suri{\'c}}, {Takalo}, {Tavecchio}, {Temnikov}, {Terzi{\'c}}, {Tescaro}, {Teshima}, {Torres}, {Toyama}, {Treves}, {Vanzo}, {Verguilov}, {Vovk}, {Ward}, {Will}, {Wu}, {Zanin}, \& {Desiante}}]{Ahnen2016_S3_HBLtpe}
---. 2016, \aap, 595, A98, \dodoi{10.1051/0004-6361/201629461}

\bibitem[{{Aleksi{\'c}} {et~al.}(2011){Aleksi{\'c}}, {Antonelli}, {Antoranz}, {Backes}, {Barrio}, {Bastieri}, {Becerra Gonz{\'a}lez}, {Bednarek}, {Berdyugin}, {Berger}, {Bernardini}, {Biland}, {Blanch}, {Bock}, {Boller}, {Bonnoli}, {Borla Tridon}, {Braun}, {Bretz}, {Ca{\~n}ellas}, {Carmona}, {Carosi}, {Colin}, {Colombo}, {Contreras}, {Cortina}, {Cossio}, {Covino}, {Dazzi}, {De Angelis}, {De Cea del Pozo}, {De Lotto}, {Delgado Mendez}, {Diago Ortega}, {Doert}, {Dom{\'\i}nguez}, {Dominis Prester}, {Dorner}, {Doro}, {Elsaesser}, {Ferenc}, {Fonseca}, {Font}, {Fruck}, {Garc{\'\i}a L{\'o}pez}, {Garczarczyk}, {Garrido}, {Giavitto}, {Godinovi{\'c}}, {Hadasch}, {H{\"a}fner}, {Herrero}, {Hildebrand}, {H{\"o}hne-M{\"o}nch}, {Hose}, {Hrupec}, {Huber}, {Jogler}, {Klepser}, {Kr{\"a}henb{\"u}hl}, {Krause}, {La Barbera}, {Lelas}, {Leonardo}, {Lindfors}, {Lombardi}, {L{\'o}pez}, {Lorenz}, {Makariev}, {Maneva}, {Mankuzhiyil}, {Mannheim}, {Maraschi}, {Mariotti}, {Mart{\'\i}nez}, {Mazin}, {Meucci}, {Miranda}, {Mirzoyan},
  {Miyamoto}, {Mold{\'o}n}, {Moralejo}, {Nieto}, {Nilsson}, {Orito}, {Oya}, {Paneque}, {Paoletti}, {Pardo}, {Paredes}, {Partini}, {Pasanen}, {Pauss}, {Perez-Torres}, {Persic}, {Peruzzo}, {Pilia}, {Pochon}, {Prada}, {Prada Moroni}, {Prandini}, {Puljak}, {Reichardt}, {Reinthal}, {Rhode}, {Rib{\'o}}, {Rico}, {R{\"u}gamer}, {Saggion}, {Saito}, {Saito}, {Salvati}, {Satalecka}, {Scalzotto}, {Scapin}, {Schultz}, {Schweizer}, {Shayduk}, {Shore}, {Sillanp{\"a}{\"a}}, {Sitarek}, {Sobczynska}, {Spanier}, {Spiro}, {Stamerra}, {Steinke}, {Storz}, {Strah}, {Suri{\'c}}, {Takalo}, {Tavecchio}, {Temnikov}, {Terzi{\'c}}, {Tescaro}, {Teshima}, {Thom}, {Tibolla}, {Torres}, {Treves}, {Vankov}, {Vogler}, {Wagner}, {Weitzel}, {Zabalza}, {Zandanel}, {Zanin}, {MAGIC Collaboration}, {Tanaka}, {Wood}, \& {Buson}}]{MAGIC2011_4c_VHE_2010}
{Aleksi{\'c}}, J., {Antonelli}, L.~A., {Antoranz}, P., {et~al.} 2011, \apjl, 730, L8, \dodoi{10.1088/2041-8205/730/1/L8}

\bibitem[{{Anderhub} {et~al.}(2009){Anderhub}, {Antonelli}, {Antoranz}, {Backes}, {Baixeras}, {Balestra}, {Barrio}, {Bartko}, {Bastieri}, {Becerra Gonz{\'a}lez}, {Becker}, {Bednarek}, {Berger}, {Bernardini}, {Biland}, {Bock}, {Bonnoli}, {Bordas}, {Borla Tridon}, {Bosch-Ramon}, {Bretz}, {Britvitch}, {Camara}, {Carmona}, {Commichau}, {Contreras}, {Cortina}, {Costado}, {Covino}, {Curtef}, {Dazzi}, {de Angelis}, {de Cea Del Pozo}, {de Los Reyes}, {de Lotto}, {de Maria}, {de Sabata}, {Delgado Mendez}, {Dominguez}, {Dorner}, {Doro}, {Elsaesser}, {Errando}, {Ferenc}, {Fern{\'a}ndez}, {Firpo}, {Fonseca}, {Font}, {Galante}, {Garc{\'\i}a L{\'o}pez}, {Garczarczyk}, {Gaug}, {Goebel}, {Hadasch}, {Hayashida}, {Herrero}, {H{\"o}hne-M{\"o}nch}, {Hose}, {Hsu}, {Huber}, {Jogler}, {Kranich}, {La Barbera}, {Laille}, {Leonardo}, {Lindfors}, {Lombardi}, {Longo}, {L{\'o}pez}, {Lorenz}, {Majumdar}, {Maneva}, {Mankuzhiyil}, {Mannheim}, {Maraschi}, {Mariotti}, {Mart{\'\i}nez}, {Mazin}, {Meucci}, {Meyer}, {Miranda}, {Mirzoyan},
  {Mold{\'o}n}, {Moles}, {Moralejo}, {Nieto}, {Nilsson}, {Ninkovic}, {Otte}, {Oya}, {Paoletti}, {Paredes}, {Pasanen}, {Pascoli}, {Pauss}, {Pegna}, {Perez-Torres}, {Persic}, {Peruzzo}, {Prada}, {Prandini}, {Puchades}, {Rhode}, {Rib{\'o}}, {Rico}, {Rissi}, {Robert}, {R{\"u}gamer}, {Saggion}, {Saito}, {Salvati}, {Sanchez-Conde}, {Sartori}, {Satalecka}, {Scalzotto}, {Scapin}, {Schweizer}, {Shayduk}, {Shinozaki}, {Shore}, {Sidro}, {Sierpowska-Bartosik}, {Sillanp{\"a}{\"a}}, {Sitarek}, {Sobczynska}, {Spanier}, {Stamerra}, {Stark}, {Takalo}, {Tavecchio}, {Temnikov}, {Tescaro}, {Teshima}, {Tluczykont}, {Torres}, {Turini}, {Vankov}, {Venturini}, {Vitale}, {Wagner}, {Wittek}, {Zabalza}, {Zandanel}, {Zanin}, {Zapatero}, {Vercellone}, {Donnarumma}, {D'Ammando}, \& {Tavani}}]{2009A&A...498...83A}
{Anderhub}, H., {Antonelli}, L.~A., {Antoranz}, P., {et~al.} 2009, \aap, 498, 83, \dodoi{10.1051/0004-6361/200811326}

\bibitem[{{Angioni} {et~al.}(2019){Angioni}, {Nesci}, {Finke}, {Buson}, \& {Ciprini}}]{Angioni2019_pks0346}
{Angioni}, R., {Nesci}, R., {Finke}, J.~D., {Buson}, S., \& {Ciprini}, S. 2019, \aap, 627, A140, \dodoi{10.1051/0004-6361/201935461}

\bibitem[{{Atwood} {et~al.}(2009){Atwood}, {Abdo}, {Ackermann}, {Althouse}, {Anderson}, {Axelsson}, {Baldini}, {Ballet}, {Band}, {Barbiellini}, {Bartelt}, {Bastieri}, {Baughman}, {Bechtol}, {B{\'e}d{\'e}r{\`e}de}, {Bellardi}, {Bellazzini}, {Berenji}, {Bignami}, {Bisello}, {Bissaldi}, {Blandford}, {Bloom}, {Bogart}, {Bonamente}, {Bonnell}, {Borgland}, {Bouvier}, {Bregeon}, {Brez}, {Brigida}, {Bruel}, {Burnett}, {Busetto}, {Caliandro}, {Cameron}, {Caraveo}, {Carius}, {Carlson}, {Casandjian}, {Cavazzuti}, {Ceccanti}, {Cecchi}, {Charles}, {Chekhtman}, {Cheung}, {Chiang}, {Chipaux}, {Cillis}, {Ciprini}, {Claus}, {Cohen-Tanugi}, {Condamoor}, {Conrad}, {Corbet}, {Corucci}, {Costamante}, {Cutini}, {Davis}, {Decotigny}, {DeKlotz}, {Dermer}, {de Angelis}, {Digel}, {do Couto e Silva}, {Drell}, {Dubois}, {Dumora}, {Edmonds}, {Fabiani}, {Farnier}, {Favuzzi}, {Flath}, {Fleury}, {Focke}, {Funk}, {Fusco}, {Gargano}, {Gasparrini}, {Gehrels}, {Gentit}, {Germani}, {Giebels}, {Giglietto}, {Giommi}, {Giordano}, {Glanzman},
  {Godfrey}, {Grenier}, {Grondin}, {Grove}, {Guillemot}, {Guiriec}, {Haller}, {Harding}, {Hart}, {Hays}, {Healey}, {Hirayama}, {Hjalmarsdotter}, {Horn}, {Hughes}, {J{\'o}hannesson}, {Johansson}, {Johnson}, {Johnson}, {Johnson}, {Johnson}, {Kamae}, {Katagiri}, {Kataoka}, {Kavelaars}, {Kawai}, {Kelly}, {Kerr}, {Klamra}, {Kn{\"o}dlseder}, {Kocian}, {Komin}, {Kuehn}, {Kuss}, {Landriu}, {Latronico}, {Lee}, {Lee}, {Lemoine-Goumard}, {Lionetto}, {Longo}, {Loparco}, {Lott}, {Lovellette}, {Lubrano}, {Madejski}, {Makeev}, {Marangelli}, {Massai}, {Mazziotta}, {McEnery}, {Menon}, {Meurer}, {Michelson}, {Minuti}, {Mirizzi}, {Mitthumsiri}, {Mizuno}, {Moiseev}, {Monte}, {Monzani}, {Moretti}, {Morselli}, {Moskalenko}, {Murgia}, {Nakamori}, {Nishino}, {Nolan}, {Norris}, {Nuss}, {Ohno}, {Ohsugi}, {Omodei}, {Orlando}, {Ormes}, {Paccagnella}, {Paneque}, {Panetta}, {Parent}, {Pearce}, {Pepe}, {Perazzo}, {Pesce-Rollins}, {Picozza}, {Pieri}, {Pinchera}, {Piron}, {Porter}, {Poupard}, {Rain{\`o}}, {Rando}, {Rapposelli}, {Razzano},
  {Reimer}, {Reimer}, {Reposeur}, {Reyes}, {Ritz}, {Rochester}, {Rodriguez}, {Romani}, {Roth}, {Russell}, {Ryde}, {Sabatini}, {Sadrozinski}, {Sanchez}, {Sander}, {Sapozhnikov}, {Parkinson}, {Scargle}, {Schalk}, {Scolieri}, {Sgr{\`o}}, {Share}, {Shaw}, {Shimokawabe}, {Shrader}, {Sierpowska-Bartosik}, {Siskind}, {Smith}, {Smith}, {Spandre}, {Spinelli}, {Starck}, {Stephens}, {Strickman}, {Strong}, {Suson}, {Tajima}, {Takahashi}, {Takahashi}, {Tanaka}, {Tenze}, {Tether}, {Thayer}, {Thayer}, {Thompson}, {Tibaldo}, {Tibolla}, {Torres}, {Tosti}, {Tramacere}, {Turri}, {Usher}, {Vilchez}, {Vitale}, {Wang}, {Watters}, {Winer}, {Wood}, {Ylinen}, \& {Ziegler}}]{FermiLAT}
{Atwood}, W.~B., {Abdo}, A.~A., {Ackermann}, M., {et~al.} 2009, \apj, 697, 1071, \dodoi{10.1088/0004-637X/697/2/1071}

\bibitem[{Ballet {et~al.}(2024)Ballet, Bruel, Burnett, Lott, \& collaboration}]{ballet2024_4fglCatA}
Ballet, J., Bruel, P., Burnett, T.~H., Lott, B., \& collaboration, T. F.-L. 2024, Fermi Large Area Telescope Fourth Source Catalog Data Release 4 (4FGL-DR4).
\newblock \doarXiv{2307.12546}

\bibitem[{{Blandford} \& {Rees}(1978)}]{Blandford_Rees1978}
{Blandford}, R.~D., \& {Rees}, M.~J. 1978, \physscr, 17, 265, \dodoi{10.1088/0031-8949/17/3/020}

\bibitem[{{B{\l}a{\.z}ejowski} {et~al.}(2000){B{\l}a{\.z}ejowski}, {Sikora}, {Moderski}, \& {Madejski}}]{EC_IR}
{B{\l}a{\.z}ejowski}, M., {Sikora}, M., {Moderski}, R., \& {Madejski}, G.~M. 2000, \apj, 545, 107, \dodoi{10.1086/317791}

\bibitem[{{Bloom} \& {Marscher}(1996)}]{SSCmodel}
{Bloom}, S.~D., \& {Marscher}, A.~P. 1996, \apj, 461, 657, \dodoi{10.1086/177092}

\bibitem[{{B{\"o}ttcher} \& {Els}(2016)}]{bottcher2016}
{B{\"o}ttcher}, M., \& {Els}, P. 2016, \apj, 821, 102, \dodoi{10.3847/0004-637X/821/2/102}

\bibitem[{{B{\"o}ttcher} {et~al.}(2013){B{\"o}ttcher}, {Reimer}, {Sweeney}, \& {Prakash}}]{Bottcher2013}
{B{\"o}ttcher}, M., {Reimer}, A., {Sweeney}, K., \& {Prakash}, A. 2013, \apj, 768, 54, \dodoi{10.1088/0004-637X/768/1/54}

\bibitem[{{Burbidge} \& {Rosenberg}(1965)}]{Burbidge1965_3C_redshift}
{Burbidge}, E.~M., \& {Rosenberg}, F.~D. 1965, \apj, 142, 1673, \dodoi{10.1086/148458}

\bibitem[{{Cerruti}(2015)}]{2015_4c_VHE_2014}
{Cerruti}, M. 2015, arXiv e-prints, arXiv:1501.03554, \dodoi{10.48550/arXiv.1501.03554}

\bibitem[{{Cohen} {et~al.}(2003){Cohen}, {Lawrence}, \& {Blandford}}]{Cohen2000S3_redshift}
{Cohen}, J.~G., {Lawrence}, C.~R., \& {Blandford}, R.~D. 2003, \apj, 583, 67, \dodoi{10.1086/344837}

\bibitem[{{Costamante} {et~al.}(2018){Costamante}, {Cutini}, {Tosti}, {Antolini}, \& {Tramacere}}]{Costamante}
{Costamante}, L., {Cutini}, S., {Tosti}, G., {Antolini}, E., \& {Tramacere}, A. 2018, \mnras, 477, 4749, \dodoi{10.1093/mnras/sty887}

\bibitem[{{Dathan} \& {Kushwaha}(2025)}]{2025arXiv250405083D}
{Dathan}, L., \& {Kushwaha}, P. 2025, arXiv e-prints, arXiv:2504.05083, \dodoi{10.48550/arXiv.2504.05083}

\bibitem[{{Dermer} \& {Schlickeiser}(1992)}]{EC_AD}
{Dermer}, C.~D., \& {Schlickeiser}, R. 1992, Science, 257, 1642, \dodoi{10.1126/science.257.5077.1642}

\bibitem[{Fan {et~al.}(2008)Fan, Yuan, Liu, Hua, Romero, Zhang, Su, Gupta, Liu, Huang, Guo, \& Zhang}]{Faan2008}
Fan, J.~H., Yuan, Y.-H., Liu, Y., {et~al.} 2008, Publications of The Astronomical Society of Japan - PUBL ASTRON SOC JPN, 60, \dodoi{10.1093/pasj/60.4.707}

\bibitem[{{Fossati} {et~al.}(1998){Fossati}, {Maraschi}, {Celotti}, {Comastri}, \& {Ghisellini}}]{blazarClassification}
{Fossati}, G., {Maraschi}, L., {Celotti}, A., {Comastri}, A., \& {Ghisellini}, G. 1998, \mnras, 299, 433, \dodoi{10.1046/j.1365-8711.1998.01828.x}

\bibitem[{{Gao} {et~al.}(2019){Gao}, {Fedynitch}, {Winter}, \& {Pohl}}]{Gao}
{Gao}, S., {Fedynitch}, A., {Winter}, W., \& {Pohl}, M. 2019, Nature Astronomy, 3, 88, \dodoi{10.1038/s41550-018-0610-1}

\bibitem[{{Ghisellini} {et~al.}(2017){Ghisellini}, {Righi}, {Costamante}, \& {Tavecchio}}]{2017MNRAS.469..255G}
{Ghisellini}, G., {Righi}, C., {Costamante}, L., \& {Tavecchio}, F. 2017, \mnras, 469, 255, \dodoi{10.1093/mnras/stx806}

\bibitem[{{Ghisellini} \& {Tavecchio}(2009)}]{Ghisellini_tavecchio2009}
{Ghisellini}, G., \& {Tavecchio}, F. 2009, \mnras, 397, 985, \dodoi{10.1111/j.1365-2966.2009.15007.x}

\bibitem[{{Gravity Collaboration} {et~al.}(2018){Gravity Collaboration}, {Sturm}, {Dexter}, {Pfuhl}, {Stock}, {Davies}, {Lutz}, {Cl{\'e}net}, {Eckart}, {Eisenhauer}, {Genzel}, {Gratadour}, {H{\"o}nig}, {Kishimoto}, {Lacour}, {Millour}, {Netzer}, {Perrin}, {Peterson}, {Petrucci}, {Rouan}, {Waisberg}, {Woillez}, {Amorim}, {Brandner}, {F{\"o}rster Schreiber}, {Garcia}, {Gillessen}, {Ott}, {Paumard}, {Perraut}, {Scheithauer}, {Straubmeier}, {Tacconi}, \& {Widmann}}]{2018Natur.563..657G}
{Gravity Collaboration}, {Sturm}, E., {Dexter}, J., {et~al.} 2018, \nat, 563, 657, \dodoi{10.1038/s41586-018-0731-9}

\bibitem[{{H.~E.~S.~S. Collaboration} {et~al.}(2010){H.~E.~S.~S. Collaboration}, {Abramowski}, {Acero}, {Aharonian}, {Akhperjanian}, {Anton}, {Barres de Almeida}, {Bazer-Bachi}, {Becherini}, {Benbow}, {Bernl{\"o}hr}, {Bochow}, {Boisson}, {Bolmont}, {Borrel}, {Brucker}, {Brun}, {Brun}, {B{\"u}hler}, {Bulik}, {B{\"u}sching}, {Boutelier}, {Chadwick}, {Charbonnier}, {Chaves}, {Cheesebrough}, {Chounet}, {Clapson}, {Coignet}, {Conrad}, {Costamante}, {Dalton}, {Daniel}, {Davids}, {Degrange}, {Deil}, {Dickinson}, {Djannati-Ata{\"\i}}, {Domainko}, {O'C. Drury}, {Dubois}, {Dubus}, {Dyks}, {Dyrda}, {Egberts}, {Eger}, {Espigat}, {Fallon}, {Farnier}, {Fegan}, {Feinstein}, {Fernandes}, {Fiasson}, {F{\"o}rster}, {Fontaine}, {F{\"u}{\ss}ling}, {Gabici}, {Gallant}, {G{\'e}rard}, {Gerbig}, {Giebels}, {Glicenstein}, {Gl{\"u}ck}, {Goret}, {G{\"o}ring}, {Hampf}, {Hauser}, {Heinz}, {Heinzelmann}, {Henri}, {Hermann}, {Hinton}, {Hoffmann}, {Hofmann}, {Hofverberg}, {Holleran}, {Hoppe}, {Horns}, {Jacholkowska}, {de Jager}, {Jahn},
  {Jung}, {Katarzy{\'n}ski}, {Katz}, {Kaufmann}, {Kerschhaggl}, {Khangulyan}, {Kh{\'e}lifi}, {Keogh}, {Klochkov}, {Klu{\'z}niak}, {Kneiske}, {Komin}, {Kosack}, {Kossakowski}, {Lamanna}, {Lenain}, {Lohse}, {Lu}, {Marandon}, {Marcowith}, {Masbou}, {Maurin}, {McComb}, {Medina}, {M{\'e}hault}, {Moderski}, {Moulin}, {Naumann-Godo}, {de Naurois}, {Nedbal}, {Nekrassov}, {Nguyen}, {Nicholas}, {Niemiec}, {Nolan}, {Ohm}, {Olive}, {de O{\~n}a Wilhelmi}, {Opitz}, {Orford}, {Ostrowski}, {Panter}, {Paz Arribas}, {Pedaletti}, {Pelletier}, {Petrucci}, {Pita}, {P{\"u}hlhofer}, {Punch}, {Quirrenbach}, {Raubenheimer}, {Raue}, {Rayner}, {Reimer}, {Renaud}, {de los Reyes}, {Rieger}, {Ripken}, {Rob}, {Rosier-Lees}, {Rowell}, {Rudak}, {Rulten}, {Ruppel}, {Ryde}, {Sahakian}, {Santangelo}, {Schlickeiser}, {Sch{\"o}ck}, {Sch{\"o}nwald}, {Schwanke}, {Schwarzburg}, {Schwemmer}, {Shalchi}, {Sushch}, {Sikora}, {Skilton}, {Sol}, {Stawarz}, {Steenkamp}, {Stegmann}, {Stinzing}, {Superina}, {Szostek}, {Tam}, {Tavernet}, {Terrier}, {Tibolla},
  {Tluczykont}, {Valerius}, {van Eldik}, {Vasileiadis}, {Venter}, {Venter}, {Vialle}, {Viana}, {Vincent}, {Vivier}, {V{\"o}lk}, {Volpe}, {Vorobiov}, {Wagner}, {Ward}, {Zdziarski}, {Zech}, \& {Zechlin}}]{2010A&A...520A..83H}
{H.~E.~S.~S. Collaboration}, {Abramowski}, A., {Acero}, F., {et~al.} 2010, \aap, 520, A83, \dodoi{10.1051/0004-6361/201014484}

\bibitem[{{H.~E.~S.~S. Collaboration} {et~al.}(2013){H.~E.~S.~S. Collaboration}, {Abramowski}, {Acero}, {Aharonian}, {Akhperjanian}, {Anton}, {Balenderan}, {Balzer}, {Barnacka}, {Becherini}, {Becker Tjus}, {Behera}, {Bernl{\"o}hr}, {Birsin}, {Biteau}, {Bochow}, {Boisson}, {Bolmont}, {Bordas}, {Brucker}, {Brun}, {Brun}, {Bulik}, {Carrigan}, {Casanova}, {Cerruti}, {Chadwick}, {Chaves}, {Cheesebrough}, {Colafrancesco}, {Cologna}, {Conrad}, {Couturier}, {Dalton}, {Daniel}, {Davids}, {Degrange}, {Deil}, {deWilt}, {Dickinson}, {Djannati-Ata{\"\i}}, {Domainko}, {O'C. Drury}, {Dubus}, {Dutson}, {Dyks}, {Dyrda}, {Egberts}, {Eger}, {Espigat}, {Fallon}, {Farnier}, {Fegan}, {Feinstein}, {Fernandes}, {Fernandez}, {Fiasson}, {Fontaine}, {F{\"o}rster}, {F{\"u}{\ss}ling}, {Gajdus}, {Gallant}, {Garrigoux}, {Gast}, {Giebels}, {Glicenstein}, {Gl{\"u}ck}, {G{\"o}ring}, {Grondin}, {Grudzi{\'n}ska}, {H{\"a}ffner}, {Hague}, {Hahn}, {Hampf}, {Harris}, {Hauser}, {Heinz}, {Heinzelmann}, {Henri}, {Hermann}, {Hillert}, {Hinton},
  {Hofmann}, {Hofverberg}, {Holler}, {Horns}, {Jacholkowska}, {Jahn}, {Jamrozy}, {Jung}, {Kastendieck}, {Katarzy{\'n}ski}, {Katz}, {Kaufmann}, {Kh{\'e}lifi}, {Klepser}, {Klochkov}, {Klu{\'z}niak}, {Kneiske}, {Kolitzus}, {Komin}, {Kosack}, {Kossakowski}, {Krayzel}, {Kr{\"u}ger}, {Laffon}, {Lamanna}, {Lefaucheur}, {Lemoine-Goumard}, {Lenain}, {Lennarz}, {Lohse}, {Lopatin}, {Lu}, {Marandon}, {Marcowith}, {Masbou}, {Maurin}, {Maxted}, {Mayer}, {McComb}, {Medina}, {M{\'e}hault}, {Menzler}, {Moderski}, {Mohamed}, {Moulin}, {Naumann}, {Naumann-Godo}, {de Naurois}, {Nedbal}, {Nguyen}, {Niemiec}, {Nolan}, {Ohm}, {de O{\~n}a Wilhelmi}, {Opitz}, {Ostrowski}, {Oya}, {Panter}, {Parsons}, {Paz Arribas}, {Pekeur}, {Pelletier}, {Perez}, {Petrucci}, {Peyaud}, {Pita}, {P{\"u}hlhofer}, {Punch}, {Quirrenbach}, {Raab}, {Raue}, {Reimer}, {Reimer}, {Renaud}, {de los Reyes}, {Rieger}, {Ripken}, {Rob}, {Rosier-Lees}, {Rowell}, {Rudak}, {Rulten}, {Sahakian}, {Sanchez}, {Santangelo}, {Schlickeiser}, {Schulz}, {Schwanke}, {Schwarzburg},
  {Schwemmer}, {Sheidaei}, {Skilton}, {Sol}, {Spengler}, {Stawarz}, {Steenkamp}, {Stegmann}, {Stinzing}, {Stycz}, {Sushch}, {Szostek}, {Tavernet}, {Terrier}, {Tluczykont}, {Trichard}, {Valerius}, {van Eldik}, {Vasileiadis}, {Venter}, {Viana}, {Vincent}, {V{\"o}lk}, {Volpe}, {Vorobiov}, {Vorster}, {Wagner}, {Ward}, \& {White}}]{HESS2013_1510_VHE_2009}
---. 2013, \aap, 554, A107, \dodoi{10.1051/0004-6361/201321135}

\bibitem[{{H.~E.~S.~S. Collaboration} {et~al.}(2019){H.~E.~S.~S. Collaboration}, {Abdalla}, {Adam}, {Aharonian}, {Ait Benkhali}, {Ang{\"u}ner}, {Arakawa}, {Arcaro}, {Armand}, {Ashkar}, {Backes}, {Barbosa Martins}, {Barnard}, {Becherini}, {Berge}, {Bernl{\"o}hr}, {Blackwell}, {B{\"o}ttcher}, {Boisson}, {Bolmont}, {Bonnefoy}, {Bregeon}, {Breuhaus}, {Brun}, {Brun}, {Bryan}, {B{\"u}chele}, {Bulik}, {Bylund}, {Capasso}, {Caroff}, {Carosi}, {Casanova}, {Cerruti}, {Chand}, {Chandra}, {Chen}, {Colafrancesco}, {Cury{\l}o}, {Davids}, {Deil}, {Devin}, {deWilt}, {Dirson}, {Djannati-Ata{\"\i}}, {Dmytriiev}, {Donath}, {Doroshenko}, {Drury}, {Dyks}, {Egberts}, {Emery}, {Ernenwein}, {Eschbach}, {Feijen}, {Fegan}, {Fiasson}, {Fontaine}, {Funk}, {F{\"u}{\ss}ling}, {Gabici}, {Gallant}, {Gat{\'e}}, {Giavitto}, {Glawion}, {Glicenstein}, {Gottschall}, {Grondin}, {Hahn}, {Haupt}, {Heinzelmann}, {Henri}, {Hermann}, {Hinton}, {Hofmann}, {Hoischen}, {Holch}, {Holler}, {Horns}, {Huber}, {Iwasaki}, {Jamrozy}, {Jankowsky}, {Jankowsky},
  {Jardin-Blicq}, {Jung-Richardt}, {Kastendieck}, {Katarzy{\'n}ski}, {Katsuragawa}, {Katz}, {Khangulyan}, {Kh{\'e}lifi}, {King}, {Klepser}, {Klu{\'z}niak}, {Komin}, {Kosack}, {Kostunin}, {Kraus}, {Lamanna}, {Lau}, {Lemi{\`e}re}, {Lemoine-Goumard}, {Lenain}, {Leser}, {Levy}, {Lohse}, {Lypova}, {Mackey}, {Majumdar}, {Malyshev}, {Marandon}, {Marcowith}, {Mares}, {Mariaud}, {Mart{\'\i}-Devesa}, {Marx}, {Maurin}, {Meintjes}, {Mitchell}, {Moderski}, {Mohamed}, {Mohrmann}, {Moore}, {Moulin}, {Muller}, {Murach}, {Nakashima}, {de Naurois}, {Ndiyavala}, {Niederwanger}, {Niemiec}, {Oakes}, {O'Brien}, {Odaka}, {Ohm}, {de Ona Wilhelmi}, {Ostrowski}, {Oya}, {Panter}, {Parsons}, {Perennes}, {Petrucci}, {Peyaud}, {Piel}, {Pita}, {Poireau}, {Priyana Noel}, {Prokhorov}, {Prokoph}, {P{\"u}hlhofer}, {Punch}, {Quirrenbach}, {Raab}, {Rauth}, {Reimer}, {Reimer}, {Remy}, {Renaud}, {Rieger}, {Rinchiuso}, {Romoli}, {Rowell}, {Rudak}, {Ruiz-Velasco}, {Sahakian}, {Saito}, {Sanchez}, {Santangelo}, {Sasaki}, {Schlickeiser},
  {Sch{\"u}ssler}, {Schulz}, {Schutte}, {Schwanke}, {Schwemmer}, {Seglar-Arroyo}, {Senniappan}, {Seyffert}, {Shafi}, {Shiningayamwe}, {Simoni}, {Sinha}, {Sol}, {Specovius}, {Spir-Jacob}, {Stawarz}, {Steenkamp}, {Stegmann}, {Steppa}, {Takahashi}, {Tavernier}, {Taylor}, {Terrier}, {Tiziani}, {Tluczykont}, {Trichard}, {Tsirou}, {Tsuji}, \& {Tuffs}}]{HESS2019_3C279_2015_VHE}
{H.~E.~S.~S. Collaboration}, {Abdalla}, H., {Adam}, R., {et~al.} 2019, \aap, 627, A159, \dodoi{10.1051/0004-6361/201935704}

\bibitem[{{H.~E.~S.~S. Collaboration} {et~al.}(2020){H.~E.~S.~S. Collaboration}, {Abdalla}, {Adam}, {Aharonian}, {Ait Benkhali}, {Ang{\"u}ner}, {Arakawa}, {Arcaro}, {Armand}, {Ashkar}, {Backes}, {Barbosa Martins}, {Barnard}, {Becherini}, {Berge}, {Bernl{\"o}hr}, {Blackwell}, {B{\"o}ttcher}, {Boisson}, {Bolmont}, {Bonnefoy}, {Bregeon}, {Breuhaus}, {Brun}, {Brun}, {Bryan}, {B{\"u}chele}, {Bulik}, {Bylund}, {Capasso}, {Caroff}, {Carosi}, {Casanova}, {Cerruti}, {Chand}, {Chandra}, {Chen}, {Colafrancesco}, {Cury{\l}o}, {Davids}, {Deil}, {Devin}, {deWilt}, {Dirson}, {Djannati-Ata{\"\i}}, {Dmytriiev}, {Donath}, {Doroshenko}, {Drury}, {Dyks}, {Egberts}, {Emery}, {Ernenwein}, {Eschbach}, {Feijen}, {Fegan}, {Fiasson}, {Fontaine}, {Funk}, {F{\"u}{\ss}ling}, {Gabici}, {Gallant}, {Gat{\'e}}, {Giavitto}, {Glawion}, {Glicenstein}, {Gottschall}, {Grondin}, {Hahn}, {Haupt}, {Heinzelmann}, {Henri}, {Hermann}, {Hinton}, {Hofmann}, {Hoischen}, {Holch}, {Holler}, {Horns}, {Huber}, {Iwasaki}, {Jamrozy}, {Jankowsky}, {Jankowsky},
  {Jardin-Blicq}, {Jung-Richardt}, {Kastendieck}, {Katarzy{\'n}ski}, {Katsuragawa}, {Katz}, {Khangulyan}, {Kh{\'e}lifi}, {King}, {Klepser}, {Klu{\'z}niak}, {Komin}, {Kosack}, {Kostunin}, {Kraus}, {Lamanna}, {Lau}, {Lemi{\`e}re}, {Lemoine-Goumard}, {Lenain}, {Leser}, {Levy}, {Lohse}, {Lypova}, {Mackey}, {Majumdar}, {Malyshev}, {Marandon}, {Marcowith}, {Mares}, {Mariaud}, {Mart{\'\i}-Devesa}, {Marx}, {Maurin}, {Meintjes}, {Mitchell}, {Moderski}, {Mohamed}, {Mohrmann}, {Muller}, {Moore}, {Moulin}, {Murach}, {Nakashima}, {de Naurois}, {Ndiyavala}, {Niederwanger}, {Niemiec}, {Oakes}, {O'Brien}, {Odaka}, {Ohm}, {de O{\~n}a Wilhelmi}, {Ostrowski}, {Oya}, {Panter}, {Parsons}, {Perennes}, {Petrucci}, {Peyaud}, {Piel}, {Pita}, {Poireau}, {Priyana Noel}, {Prokhorov}, {Prokoph}, {P{\"u}hlhofer}, {Punch}, {Quirrenbach}, {Raab}, {Rauth}, {Reimer}, {Reimer}, {Remy}, {Renaud}, {Rieger}, {Rinchiuso}, {Romoli}, {Rowell}, {Rudak}, {Ruiz-Velasco}, {Sahakian}, {Saito}, {Sanchez}, {Santangelo}, {Sasaki}, {Schlickeiser},
  {Sch{\"u}ssler}, {Schulz}, {Schutte}, {Schwanke}, {Schwemmer}, {Seglar-Arroyo}, {Senniappan}, {Seyffert}, {Shafi}, {Shiningayamwe}, {Simoni}, {Sinha}, {Sol}, {Specovius}, {Spir-Jacob}, {Stawarz}, {Steenkamp}, {Stegmann}, {Steppa}, {Takahashi}, {Tavernier}, {Taylor}, {Terrier}, {Tiziani}, {Tluczykont}, {Trichard}, {Tsirou}, {Tsuji}, \& {Tuffs}}]{HESS2020_0736_VHE}
---. 2020, \aap, 633, A162, \dodoi{10.1051/0004-6361/201935906}

\bibitem[{{Hayashida} {et~al.}(2015){Hayashida}, {Nalewajko}, {Madejski}, {Sikora}, {Itoh}, {Ajello}, {Blandford}, {Buson}, {Chiang}, {Fukazawa}, {Furniss}, {Urry}, {Hasan}, {Harrison}, {Alexander}, {Balokovi{\'c}}, {Barret}, {Boggs}, {Christensen}, {Craig}, {Forster}, {Giommi}, {Grefenstette}, {Hailey}, {Hornstrup}, {Kitaguchi}, {Koglin}, {Madsen}, {Mao}, {Miyasaka}, {Mori}, {Perri}, {Pivovaroff}, {Puccetti}, {Rana}, {Stern}, {Tagliaferri}, {Westergaard}, {Zhang}, {Zoglauer}, {Gurwell}, {Uemura}, {Akitaya}, {Kawabata}, {Kawaguchi}, {Kanda}, {Moritani}, {Takaki}, {Ui}, {Yoshida}, {Agarwal}, \& {Gupta}}]{Hayashida2015}
{Hayashida}, M., {Nalewajko}, K., {Madejski}, G.~M., {et~al.} 2015, \apj, 807, 79, \dodoi{10.1088/0004-637X/807/1/79}

\bibitem[{{Hewett} \& {Wild}(2010)}]{Hewett2010TON_B2_redshift}
{Hewett}, P.~C., \& {Wild}, V. 2010, \mnras, 405, 2302, \dodoi{10.1111/j.1365-2966.2010.16648.x}

\bibitem[{{Ho} \& {Kim}(2009)}]{Ho2009PKS0736_redshift}
{Ho}, L.~C., \& {Kim}, M. 2009, \apjs, 184, 398, \dodoi{10.1088/0067-0049/184/2/398}

\bibitem[{{Itoh} {et~al.}(2016){Itoh}, {Nalewajko}, {Fukazawa}, {Uemura}, {Tanaka}, {Kawabata}, {Madejski}, {Schinzel}, {Kanda}, {Shiki}, {Akitaya}, {Kawabata}, {Moritani}, {Nakaoka}, {Ohsugi}, {Sasada}, {Takaki}, {Takata}, {Ui}, {Yamanaka}, \& {Yoshida}}]{Itoh2016}
{Itoh}, R., {Nalewajko}, K., {Fukazawa}, Y., {et~al.} 2016, \apj, 833, 77, \dodoi{10.3847/1538-4357/833/1/77}

\bibitem[{{Jaffe} {et~al.}(2004){Jaffe}, {Meisenheimer}, {R{\"o}ttgering}, {Leinert}, {Richichi}, {Chesneau}, {Fraix-Burnet}, {Glazenborg-Kluttig}, {Granato}, {Graser}, {Heijligers}, {K{\"o}hler}, {Malbet}, {Miley}, {Paresce}, {Pel}, {Perrin}, {Przygodda}, {Schoeller}, {Sol}, {Waters}, {Weigelt}, {Woillez}, \& {de Zeeuw}}]{2004Natur.429...47J}
{Jaffe}, W., {Meisenheimer}, K., {R{\"o}ttgering}, H.~J.~A., {et~al.} 2004, \nat, 429, 47, \dodoi{10.1038/nature02531}

\bibitem[{Knuth(2013)}]{knuth2013optimal}
Knuth, K.~H. 2013, Optimal Data-Based Binning for Histograms.
\newblock \doarXiv{physics/0605197}

\bibitem[{Kushwaha(2025)}]{universe11030084}
Kushwaha, P. 2025, Universe, 11, \dodoi{10.3390/universe11030084}

\bibitem[{{Kushwaha} {et~al.}(2016){Kushwaha}, {Chandra}, {Misra}, {Sahayanathan}, {Singh}, \& {Baliyan}}]{2016ApJ...822L..13K}
{Kushwaha}, P., {Chandra}, S., {Misra}, R., {et~al.} 2016, \apjl, 822, L13, \dodoi{10.3847/2041-8205/822/1/L13}

\bibitem[{{Kushwaha} {et~al.}(2021){Kushwaha}, {Pal}, {Kalita}, {Kumari}, {Naik}, {Gupta}, {de Gouveia Dal Pino}, \& {Gu}}]{2021ApJ...921...18K}
{Kushwaha}, P., {Pal}, M., {Kalita}, N., {et~al.} 2021, \apj, 921, 18, \dodoi{10.3847/1538-4357/ac19b8}

\bibitem[{{Kushwaha} {et~al.}(2017){Kushwaha}, {Sinha}, {Misra}, {Singh}, \& {de Gouveia Dal Pino}}]{2017ApJ...849..138K}
{Kushwaha}, P., {Sinha}, A., {Misra}, R., {Singh}, K.~P., \& {de Gouveia Dal Pino}, E.~M. 2017, \apj, 849, 138, \dodoi{10.3847/1538-4357/aa8ef5}

\bibitem[{{Kushwaha} {et~al.}(2018){Kushwaha}, {Gupta}, {Wiita}, {Pal}, {Gaur}, {de Gouveia Dal Pino}, {Kurtanidze}, {Semkov}, {Damljanovic}, {Hu}, {Uemura}, {Vince}, {Darriba}, {Gu}, {Bachev}, {Chen}, {Itoh}, {Kawabata}, {Kurtanidze}, {Nakaoka}, {Nikolashvili}, {Sigua}, {Strigachev}, \& {Zhang}}]{2018MNRAS.479.1672K}
{Kushwaha}, P., {Gupta}, A.~C., {Wiita}, P.~J., {et~al.} 2018, \mnras, 479, 1672, \dodoi{10.1093/mnras/sty1499}

\bibitem[{{Liodakis} \& {Petropoulou}(2020)}]{Liodakis_2020}
{Liodakis}, I., \& {Petropoulou}, M. 2020, \apjl, 893, L20, \dodoi{10.3847/2041-8213/ab830a}

\bibitem[{{MAGIC Collaboration} {et~al.}(2018){MAGIC Collaboration}, {Acciari}, {Ansoldi}, {Antonelli}, {Arbet Engels}, {Arcaro}, {Baack}, {Babi{\'c}}, {Banerjee}, {Bangale}, {Barres de Almeida}, {Barrio}, {Bednarek}, {Bernardini}, {Berti}, {Besenrieder}, {Bhattacharyya}, {Bigongiari}, {Biland}, {Blanch}, {Bonnoli}, {Carosi}, {Ceribella}, {Cikota}, {Colak}, {Colin}, {Colombo}, {Contreras}, {Cortina}, {Covino}, {D'Elia}, {da Vela}, {Dazzi}, {de Angelis}, {de Lotto}, {Delfino}, {Delgado}, {di Pierro}, {Do Souto Espi{\~n}era}, {Dom{\'\i}nguez}, {Dominis Prester}, {Dorner}, {Doro}, {Einecke}, {Elsaesser}, {Fallah Ramazani}, {Fattorini}, {Fern{\'a}ndez-Barral}, {Ferrara}, {Fidalgo}, {Foffano}, {Fonseca}, {Font}, {Fruck}, {Galindo}, {Gallozzi}, {Garc{\'\i}a L{\'o}pez}, {Garczarczyk}, {Gaug}, {Giammaria}, {Godinovi{\'c}}, {Guberman}, {Hadasch}, {Hahn}, {Hassan}, {Herrera}, {Hoang}, {Hrupec}, {Inoue}, {Ishio}, {Iwamura}, {Kubo}, {Kushida}, {Kuve{\v{z}}di{\'c}}, {Lamastra}, {Lelas}, {Leone}, {Lindfors}, {Lombardi},
  {Longo}, {L{\'o}pez}, {L{\'o}pez-Oramas}, {Maggio}, {Majumdar}, {Makariev}, {Maneva}, {Manganaro}, {Mannheim}, {Maraschi}, {Mariotti}, {Mart{\'\i}nez}, {Masuda}, {Mazin}, {Minev}, {Miranda}, {Mirzoyan}, {Molina}, {Moralejo}, {Moreno}, {Moretti}, {Munar-Adrover}, {Neustroev}, {Niedzwiecki}, {Nievas Rosillo}, {Nigro}, {Nilsson}, {Ninci}, {Nishijima}, {Noda}, {Nogu{\'e}s}, {Paiano}, {Palacio}, {Paneque}, {Paoletti}, {Paredes}, {Pedaletti}, {Pe{\~n}il}, {Peresano}, {Persic}, {Prada Moroni}, {Prandini}, {Puljak}, {Garcia}, {Rhode}, {Rib{\'o}}, {Rico}, {Righi}, {Rugliancich}, {Saha}, {Saito}, {Satalecka}, {Schweizer}, {Sitarek}, {{\v{S}}nidari{\'c}}, {Sobczynska}, {Somero}, {Stamerra}, {Strzys}, {Suri{\'c}}, {Tavecchio}, {Temnikov}, {Terzi{\'c}}, {Teshima}, {Torres-Alb{\`a}}, {Tsujimoto}, {van Scherpenberg}, {Vanzo}, {Vazquez Acosta}, {Vovk}, {Ward}, {Will}, {Zari{\'c}}, {Fermi-Lat Collaboration}, {Becerra Gonz{\'a}lez}, {Raiteri}, {Sandrinelli}, {Hovatta}, {Kiehlmann}, {Max-Moerbeck}, {Tornikoski},
  {L{\"a}hteenm{\"a}ki}, {Tammi}, {Ramakrishnan}, {Thum}, {Agudo}, {Molina}, {G{\'o}mez}, {Fuentes}, {Casadio}, {Traianou}, {Myserlis}, \& {Kim}}]{2018Magic_PKS1510-089}
{MAGIC Collaboration}, {Acciari}, V.~A., {Ansoldi}, S., {et~al.} 2018, \aap, 619, A159, \dodoi{10.1051/0004-6361/201833618}

\bibitem[{{MAGIC Collaboration} {et~al.}(2021){MAGIC Collaboration}, {Acciari}, {Ansoldi}, {Antonelli}, {Arbet Engels}, {Artero}, {Asano}, {Baack}, {Babi{\'c}}, {Baquero}, {Barres de Almeida}, {Barrio}, {Becerra Gonz{\'a}lez}, {Bednarek}, {Bellizzi}, {Bernardini}, {Bernardos}, {Berti}, {Besenrieder}, {Bhattacharyya}, {Bigongiari}, {Biland}, {Blanch}, {Bonnoli}, {Bo{\v{s}}njak}, {Busetto}, {Carosi}, {Ceribella}, {Cerruti}, {Chai}, {Chilingarian}, {Cikota}, {Colak}, {Colombo}, {Contreras}, {Cortina}, {Covino}, {D'Amico}, {D'Elia}, {da Vela}, {Dazzi}, {de Angelis}, {de Lotto}, {Delfino}, {Delgado}, {Delgado Mendez}, {Depaoli}, {di Pierro}, {di Venere}, {Do Souto Espi{\~n}eira}, {Dominis Prester}, {Donini}, {Dorner}, {Doro}, {Elsaesser}, {Fallah Ramazani}, {Fattorini}, {Ferrara}, {Foffano}, {Fonseca}, {Font}, {Fruck}, {Fukami}, {Garc{\'\i}a L{\'o}pez}, {Garczarczyk}, {Gasparyan}, {Gaug}, {Giglietto}, {Giordano}, {Gliwny}, {Godinovi{\'c}}, {Green}, {Green}, {Hadasch}, {Hahn}, {Heckmann}, {Herrera}, {Hoang},
  {Hrupec}, {H{\"u}tten}, {Inada}, {Inoue}, {Ishio}, {Iwamura}, {Jormanainen}, {Jouvin}, {Kajiwara}, {Karjalainen}, {Kerszberg}, {Kobayashi}, {Kubo}, {Kushida}, {Lamastra}, {Lelas}, {Leone}, {Lindfors}, {Lombardi}, {Longo}, {L{\'o}pez-Coto}, {L{\'o}pez-Moya}, {L{\'o}pez-Oramas}, {Loporchio}, {Machado de Oliveira Fraga}, {Maggio}, {Majumdar}, {Makariev}, {Mallamaci}, {Maneva}, {Manganaro}, {Mannheim}, {Maraschi}, {Mariotti}, {Mart{\'\i}nez}, {Mazin}, {Mender}, {Mi{\'c}anovi{\'c}}, {Miceli}, {Miener}, {Minev}, {Miranda}, {Mirzoyan}, {Molina}, {Moralejo}, {Morcuende}, {Moreno}, {Moretti}, {Neustroev}, {Nigro}, {Nilsson}, {Ninci}, {Nishijima}, {Noda}, {Nozaki}, {Ohtani}, {Oka}, {Otero-Santos}, {Paiano}, {Palatiello}, {Paneque}, {Paoletti}, {Paredes}, {Pavleti{\'c}}, {Pe{\~n}il}, {Perennes}, {Persic}, {Prada Moroni}, {Prandini}, {Priyadarshi}, {Puljak}, {Rhode}, {Rib{\'o}}, {Rico}, {Righi}, {Rugliancich}, {Saha}, {Sahakyan}, {Saito}, {Sakurai}, {Satalecka}, {Saturni}, {Schleicher}, {Schmidt}, {Schweizer},
  {Sitarek}, {{\v{S}}nidari{\'c}}, {Sobczynska}, {Spolon}, {Stamerra}, {Strom}, {Strzys}, {Suda}, {Suri{\'c}}, {Takahashi}, {Tavecchio}, {Temnikov}, {Terzi{\'c}}, {Teshima}, {Torres-Alb{\`a}}, {Tosti}, {Truzzi}, {Tutone}, {van Scherpenberg}, {Vanzo}, {Vazquez Acosta}, {Ventura}, {Verguilov}, {Vigorito}, {Vitale}, {Vovk}, {Will}, {Zari{\'c}}, {Angioni}, {D'Ammando}, {Ciprini}, {Cheung}, {Orienti}, {Pacciani}, {Prajapati}, {Kumar}, \& {Ganesh}}]{MAGIC2021_B2_2020_VHE}
---. 2021, \aap, 647, A163, \dodoi{10.1051/0004-6361/202039687}

\bibitem[{{Malik} {et~al.}(2025){Malik}, {Akbar}, {Shah}, {Misra}, {Dar}, {Manzoor}, {Ahanger}, {Nazir}, {Iqbal}, {Rubab}, \& {Tantry}}]{2025MNRAS.539.2185M}
{Malik}, Z., {Akbar}, S., {Shah}, Z., {et~al.} 2025, \mnras, 539, 2185, \dodoi{10.1093/mnras/staf620}

\bibitem[{{Malmrose} {et~al.}(2011){Malmrose}, {Marscher}, {Jorstad}, {Nikutta}, \& {Elitzur}}]{Malmarose_2011}
{Malmrose}, M.~P., {Marscher}, A.~P., {Jorstad}, S.~G., {Nikutta}, R., \& {Elitzur}, M. 2011, \apj, 732, 116, \dodoi{10.1088/0004-637X/732/2/116}

\bibitem[{{Mannheim} \& {Biermann}(1992)}]{Mannheim1992}
{Mannheim}, K., \& {Biermann}, P.~L. 1992, \aap, 253, L21

\bibitem[{{Manzoor} {et~al.}(2024){Manzoor}, {Shah}, {Sahayanathan}, {Iqbal}, \& {Dar}}]{2024MNRAS.529.1356M}
{Manzoor}, A., {Shah}, Z., {Sahayanathan}, S., {Iqbal}, N., \& {Dar}, A.~A. 2024, \mnras, 529, 1356, \dodoi{10.1093/mnras/stae588}

\bibitem[{{Maraschi} {et~al.}(1992){Maraschi}, {Ghisellini}, \& {Celotti}}]{ICscattering}
{Maraschi}, L., {Ghisellini}, G., \& {Celotti}, A. 1992, \apjl, 397, L5, \dodoi{10.1086/186531}

\bibitem[{{Maurya} {et~al.}(2025){Maurya}, {Majumdar}, {Varun}, {Sahu}, \& {Prince}}]{2025Ton0599_rajPrince}
{Maurya}, S., {Majumdar}, J., {Varun}, {Sahu}, N., \& {Prince}, R. 2025, \pasa, 42, e053, \dodoi{10.1017/pasa.2025.34}

\bibitem[{{Meyer} \& {Georganopoulos}(2014)}]{Meyer_georganopolous2014}
{Meyer}, E.~T., \& {Georganopoulos}, M. 2014, \apjl, 780, L27, \dodoi{10.1088/2041-8205/780/2/L27}

\bibitem[{{Meyer} {et~al.}(2019){Meyer}, {Scargle}, \& {Blandford}}]{Meyer}
{Meyer}, M., {Scargle}, J.~D., \& {Blandford}, R.~D. 2019, \apj, 877, 39, \dodoi{10.3847/1538-4357/ab1651}

\bibitem[{{Middei} {et~al.}(2023){Middei}, {Liodakis}, {Perri}, {Puccetti}, {Cavazzuti}, {Di Gesu}, {Ehlert}, {Madejski}, {Marscher}, {Marshall}, {Muleri}, {Negro}, {Jorstad}, {Ag{\'\i}s-Gonz{\'a}lez}, {Agudo}, {Bonnoli}, {Bernardos}, {Casanova}, {Garc{\'\i}a-Comas}, {Husillos}, {Marchini}, {Sota}, {Kouch}, {Lindfors}, {Borman}, {Kopatskaya}, {Larionova}, {Morozova}, {Savchenko}, {Vasilyev}, {Zhovtan}, {Casadio}, {Escudero}, {Myserlis}, {Hales}, {Kameno}, {Kneissl}, {Messias}, {Nagai}, {Blinov}, {Bourbah}, {Kiehlmann}, {Kontopodis}, {Mandarakas}, {Romanopoulos}, {Skalidis}, {Vervelaki}, {Masiero}, {Mawet}, {Millar-Blanchaer}, {Panopoulou}, {Tinyanont}, {Berdyugin}, {Kagitani}, {Kravtsov}, {Sakanoi}, {Imazawa}, {Sasada}, {Fukazawa}, {Kawabata}, {Uemura}, {Mizuno}, {Nakaoka}, {Akitaya}, {Gurwell}, {Rao}, {Di Lalla}, {Cibrario}, {Donnarumma}, {Kim}, {Omodei}, {Pacciani}, {Poutanen}, {Tavecchio}, {Antonelli}, {Bachetti}, {Baldini}, {Baumgartner}, {Bellazzini}, {Bianchi}, {Bongiorno}, {Bonino}, {Brez},
  {Bucciantini}, {Capitanio}, {Castellano}, {Ciprini}, {Costa}, {De Rosa}, {Del Monte}, {Di Marco}, {Doroshenko}, {Dov{\v{c}}iak}, {Enoto}, {Evangelista}, {Fabiani}, {Ferrazzoli}, {Garcia}, {Gunji}, {Hayashida}, {Heyl}, {Iwakiri}, {Karas}, {Kitaguchi}, {Kolodziejczak}, {Krawczynski}, {La Monaca}, {Latronico}, {Maldera}, {Manfreda}, {Marin}, {Marinucci}, {Massaro}, {Matt}, {Mitsuishi}, {Ng}, {O'Dell}, {Oppedisano}, {Papitto}, {Pavlov}, {Peirson}, {Pesce-Rollins}, {Petrucci}, {Pilia}, {Possenti}, {Ramsey}, {Rankin}, {Ratheesh}, {Romani}, {Sgr{\'o}}, {Slane}, {Soffitta}, {Spandre}, {Tamagawa}, {Taverna}, {Tawara}, {Tennant}, {Thomas}, {Tombesi}, {Trois}, {Tsygankov}, {Turolla}, {Vink}, {Weisskopf}, {Wu}, {Xie}, \& {Zane}}]{Middei2023}
{Middei}, R., {Liodakis}, I., {Perri}, M., {et~al.} 2023, \apjl, 942, L10, \dodoi{10.3847/2041-8213/aca281}

\bibitem[{{Mirzoyan}(2017)}]{2017ATel_TON_2017}
{Mirzoyan}, R. 2017, The Astronomer's Telegram, 11061, 1

\bibitem[{{M{\"u}cke} \& {Protheroe}(2001)}]{Mucke2001}
{M{\"u}cke}, A., \& {Protheroe}, R.~J. 2001, Astroparticle Physics, 15, 121, \dodoi{10.1016/S0927-6505(00)00141-9}

\bibitem[{{Nievas Rosillo} {et~al.}(2024){Nievas Rosillo}, {Acero}, {Otero-Santos}, {Vazquez Acosta}, {Terrier}, {Morcuende}, \& {Arbet-Engels}}]{2024_op313_vhe}
{Nievas Rosillo}, M., {Acero}, F., {Otero-Santos}, J., {et~al.} 2024, arXiv e-prints, arXiv:2409.20487, \dodoi{10.48550/arXiv.2409.20487}

\bibitem[{{Osterbrock} \& {Pogge}(1987)}]{Osterbrock1987_4C_redshift}
{Osterbrock}, D.~E., \& {Pogge}, R.~W. 1987, \apj, 323, 108, \dodoi{10.1086/165810}

\bibitem[{{Pandey} {et~al.}(2024){Pandey}, {Kushwaha}, {Wiita}, {Prince}, {Czerny}, \& {Stalin}}]{2024A&A...681A.116P}
{Pandey}, A., {Kushwaha}, P., {Wiita}, P.~J., {et~al.} 2024, \aap, 681, A116, \dodoi{10.1051/0004-6361/202347719}

\bibitem[{{Pian} {et~al.}(1998){Pian}, {Vacanti}, {Tagliaferri}, {Ghisellini}, {Maraschi}, {Treves}, {Urry}, {Fiore}, {Giommi}, {Palazzi}, {Chiappetti}, \& {Sambruna}}]{Pian1998}
{Pian}, E., {Vacanti}, G., {Tagliaferri}, G., {et~al.} 1998, \apjl, 492, L17, \dodoi{10.1086/311083}

\bibitem[{{Roy} {et~al.}(2021){Roy}, {Patel}, {Sarkar}, {Chatterjee}, \& {Chitnis}}]{Roy2021}
{Roy}, A., {Patel}, S.~R., {Sarkar}, A., {Chatterjee}, A., \& {Chitnis}, V.~R. 2021, \mnras, 504, 1103, \dodoi{10.1093/mnras/stab975}

\bibitem[{{Sahayanathan} \& {Godambe}(2012)}]{2012MNRAS.419.1660S}
{Sahayanathan}, S., \& {Godambe}, S. 2012, \mnras, 419, 1660, \dodoi{10.1111/j.1365-2966.2011.19829.x}

\bibitem[{{Scargle} {et~al.}(2013){Scargle}, {Norris}, {Jackson}, \& {Chiang}}]{Scargle2013}
{Scargle}, J.~D., {Norris}, J.~P., {Jackson}, B., \& {Chiang}, J. 2013, \apj, 764, 167, \dodoi{10.1088/0004-637X/764/2/167}

\bibitem[{{Schneider} {et~al.}(2010){Schneider}, {Richards}, {Hall}, {Strauss}, {Anderson}, {Boroson}, {Ross}, {Shen}, {Brandt}, {Fan}, {Inada}, {Jester}, {Knapp}, {Krawczyk}, {Thakar}, {Vanden Berk}, {Voges}, {Yanny}, {York}, {Bahcall}, {Bizyaev}, {Blanton}, {Brewington}, {Brinkmann}, {Eisenstein}, {Frieman}, {Fukugita}, {Gray}, {Gunn}, {Hibon}, {Ivezi{\'c}}, {Kent}, {Kron}, {Lee}, {Lupton}, {Malanushenko}, {Malanushenko}, {Oravetz}, {Pan}, {Pier}, {Price}, {Saxe}, {Schlegel}, {Simmons}, {Snedden}, {SubbaRao}, {Szalay}, \& {Weinberg}}]{2010AJ_op313_redshift}
{Schneider}, D.~P., {Richards}, G.~T., {Hall}, P.~B., {et~al.} 2010, \aj, 139, 2360, \dodoi{10.1088/0004-6256/139/6/2360}

\bibitem[{{Shah} {et~al.}(2018){Shah}, {Mankuzhiyil}, {Sinha}, {Misra}, {Sahayanathan}, \& {Iqbal}}]{2018RAA....18..141S}
{Shah}, Z., {Mankuzhiyil}, N., {Sinha}, A., {et~al.} 2018, Research in Astronomy and Astrophysics, 18, 141, \dodoi{10.1088/1674-4527/18/11/141}

\bibitem[{{Shaw} {et~al.}(2012){Shaw}, {Romani}, {Cotter}, {Healey}, {Michelson}, {Readhead}, {Richards}, {Max-Moerbeck}, {King}, \& {Potter}}]{Shaw2012PKS1441_redshift}
{Shaw}, M.~S., {Romani}, R.~W., {Cotter}, G., {et~al.} 2012, \apj, 748, 49, \dodoi{10.1088/0004-637X/748/1/49}

\bibitem[{{Sikora} {et~al.}(1994){Sikora}, {Begelman}, \& {Rees}}]{EC_BLR}
{Sikora}, M., {Begelman}, M.~C., \& {Rees}, M.~J. 1994, \apj, 421, 153, \dodoi{10.1086/173633}

\bibitem[{{Tanner} {et~al.}(1996){Tanner}, {Bechtold}, {Walker}, {Black}, \& {Cutri}}]{Tanner1996PKS1510_redshift}
{Tanner}, A.~M., {Bechtold}, J., {Walker}, C.~E., {Black}, J.~H., \& {Cutri}, R.~M. 1996, \aj, 112, 62, \dodoi{10.1086/117988}

\bibitem[{{Tavecchio} {et~al.}(2000){Tavecchio}, {Maraschi}, {Sambruna}, \& {Urry}}]{EC_CMBR}
{Tavecchio}, F., {Maraschi}, L., {Sambruna}, R.~M., \& {Urry}, C.~M. 2000, \apjl, 544, L23, \dodoi{10.1086/317292}

\bibitem[{{Urry} \& {Padovani}(1995)}]{UrryPadovini1995}
{Urry}, C.~M., \& {Padovani}, P. 1995, \pasp, 107, 803, \dodoi{10.1086/133630}

\bibitem[{{Valverde} {et~al.}(2020){Valverde}, {Horan}, {Bernard}, {Fegan}, {Fermi-LAT Collaboration}, {Abeysekara}, {Archer}, {Benbow}, {Bird}, {Brill}, {Brose}, {Buchovecky}, {Buckley}, {Christiansen}, {Cui}, {Falcone}, {Feng}, {Finley}, {Fortson}, {Furniss}, {Gent}, {Gillanders}, {Giuri}, {Gueta}, {Hanna}, {Hassan}, {Hervet}, {Holder}, {Hughes}, {Humensky}, {Kaaret}, {Kelley-Hoskins}, {Kertzman}, {Kieda}, {Krause}, {Krennrich}, {Lang}, {Maier}, {Moriarty}, {Mukherjee}, {Nieto}, {Nievas-Rosillo}, {O'Brien}, {Ong}, {Otte}, {Park}, {Petrashyk}, {Pfrang}, {Pichel}, {Pohl}, {Prado}, {Pueschel}, {Quinn}, {Ragan}, {Reynolds}, {Ribeiro}, {Richards}, {Roache}, {Sadeh}, {Santander}, {Scott}, {Sembroski}, {Shahinyan}, {Shang}, {Sushch}, {Vassiliev}, {Weinstein}, {Wells}, {Wilcox}, {Wilhelm}, {Williams}, {Williamson}, {VERITAS Collaboration}, {Noto}, {Edwards}, {Piner}, {Fallah Ramazani}, {Hovatta}, {Jormanainen}, {Lindfors}, {Nilsson}, {Takalo}, {Kovalev}, {Lister}, {Pushkarev}, {Savolainen}, {Kiehlmann},
  {Max-Moerbeck}, {Readhead}, {L{\"a}hteenm{\"a}ki}, \& {Tornikoski}}]{2020ApJ...891..170V}
{Valverde}, J., {Horan}, D., {Bernard}, D., {et~al.} 2020, \apj, 891, 170, \dodoi{10.3847/1538-4357/ab765d}

\bibitem[{{Wagner} {et~al.}(2021){Wagner}, {Rani}, \& {H.~E.~S.~S. Collaboration}}]{2021ATel_0346_VHE}
{Wagner}, S., {Rani}, B., \& {H.~E.~S.~S. Collaboration}. 2021, The Astronomer's Telegram, 15020, 1

\bibitem[{Wagner \& Witzel(1995)}]{Wagner1995_intradayVariability}
Wagner, S.~J., \& Witzel, A. 1995, Annual Review of Astronomy and Astrophysics, 33, 163, \dodoi{https://doi.org/10.1146/annurev.aa.33.090195.001115}

\bibitem[{{Wang} {et~al.}(2023){Wang}, {Yi}, {Wang}, {Mao}, {Pu}, {Ning}, {Huang}, {Lu}, {Zhang}, {Chen}, \& {Dong}}]{2023RAA....23k5011W}
{Wang}, N., {Yi}, T.-F., {Wang}, L., {et~al.} 2023, Research in Astronomy and Astrophysics, 23, 115011, \dodoi{10.1088/1674-4527/ace9b1}

\bibitem[{{White} {et~al.}(1988){White}, {Jauncey}, {Savage}, {Wright}, {Batty}, {Peterson}, \& {Gulkis}}]{White1988PKS0346_redshift}
{White}, G.~L., {Jauncey}, D.~L., {Savage}, A., {et~al.} 1988, \apj, 327, 561, \dodoi{10.1086/166216}

\bibitem[{Yadav \& Kushwaha(2024)}]{Sameer_2024}
Yadav, S., \& Kushwaha, P. 2024, Galaxies, 12, 34, \dodoi{10.3390/galaxies12040034}

\bibitem[{{Zacharias} {et~al.}(2017){Zacharias}, {Sitarek}, {Dominis Prester}, {Jankowsky}, {Lindfors}, {Mohamed}, {Sanchez}, {Terzic}, {H.~E.~S.~S. Collaboration}, \& {MAGIC Collaboration}}]{2017_1510_VHE_2016}
{Zacharias}, M., {Sitarek}, J., {Dominis Prester}, D., {et~al.} 2017, in International Cosmic Ray Conference, Vol. 301, 35th International Cosmic Ray Conference (ICRC2017), 655, \dodoi{10.22323/1.301.0655}

\end{thebibliography}
\bibliographystyle{aasjournal}



\end{document}